\documentclass[conference, 10pt]{IEEEtran}
\usepackage{amssymb}
\usepackage{amsmath}
\usepackage{microtype}
\usepackage{graphicx}
\usepackage[numbers]{natbib}
\usepackage[utf8]{inputenc}
\usepackage[T1]{fontenc}
\usepackage{stfloats}
\usepackage{url}
\usepackage{siunitx}
\usepackage{soul}
\usepackage{color}
\usepackage{enumitem}
\usepackage{comment}
\usepackage{verbatim}
\usepackage{longtable}
\usepackage{booktabs}
\usepackage{array}
\usepackage{tabularx}
\usepackage{gensymb}
\usepackage[table, xcdraw]{xcolor}
\usepackage[normalem]{ulem}
\usepackage{threeparttable}
\usepackage{placeins}
\useunder{\uline}{\ul}{}

\ifCLASSINFOpdf
\else
\fi
\begin{document}
%
%
\title{GaN Power Devices and Converter Architectures for AI Data Centers: Efficiency, Reliability, and Deployment Pathways}

\author{\IEEEauthorblockN{Donald Intal, Abasifreke Ebong}
\IEEEauthorblockA{Department of Electrical and Computer Engineering\\
University of North Carolina at Charlotte\\
Charlotte, North Carolina 28223\\
Email: dintal@charlotte.edu, aebong1@charlotte.edu}}


%


\maketitle
\pagestyle{plain}
\pagenumbering{gobble}


%
\IEEEpeerreviewmaketitle

\begin{abstract}
The rapid growth of artificial-intelligence workloads is increasing the electrical and thermal demands placed on data-center power-delivery systems, making conversion efficiency, power density, and reliability critical infrastructure-level considerations. This review examines how gallium-nitride (GaN) power devices can be matched to specific stages of the grid-to-load conversion chain, including front-end power-factor correction, isolated DC/DC conversion, 48-V intermediate-bus conversion, and point-of-load regulation. The material and device characteristics of Si, SiC, and GaN are first compared using converter-relevant metrics, after which lateral, vertical, and specialized GaN architectures are evaluated in terms of voltage scalability, switching behavior, reverse conduction, thermal pathways, gate control, and technology maturity. The analysis shows that GaN provides a stage-dependent rather than universal advantage. Commercially mature lateral GaN HEMTs are particularly effective in high-frequency, low-to-mid-voltage stages where switching and commutation losses strongly influence efficiency and passive-component volume. Specialized and hybrid devices extend this capability to bidirectional operation, normally-off control, extreme conversion ratios, and functional integration, while vertical GaN remains an emerging option for higher-voltage and higher-power conversion. A quantitative framework is also presented to connect cascaded converter efficiency with electrical-loss reduction, cooling demand, annual facility energy use, and operational carbon emissions. The resulting benefits depend on the fraction of load processed, operating profile, cooling-system performance, and grid carbon intensity. Broad deployment further requires low-parasitic packaging, disciplined gate-drive and EMI co-design, mission-profile reliability qualification, scalable manufacturing, and supply-chain resilience. GaN is therefore best treated as a stage-specific system lever whose value emerges through coordinated device--topology--package--thermal co-design.
\end{abstract}

\section{Introduction}

\begin{figure}[!h]
\centering
\includegraphics[width=\columnwidth]{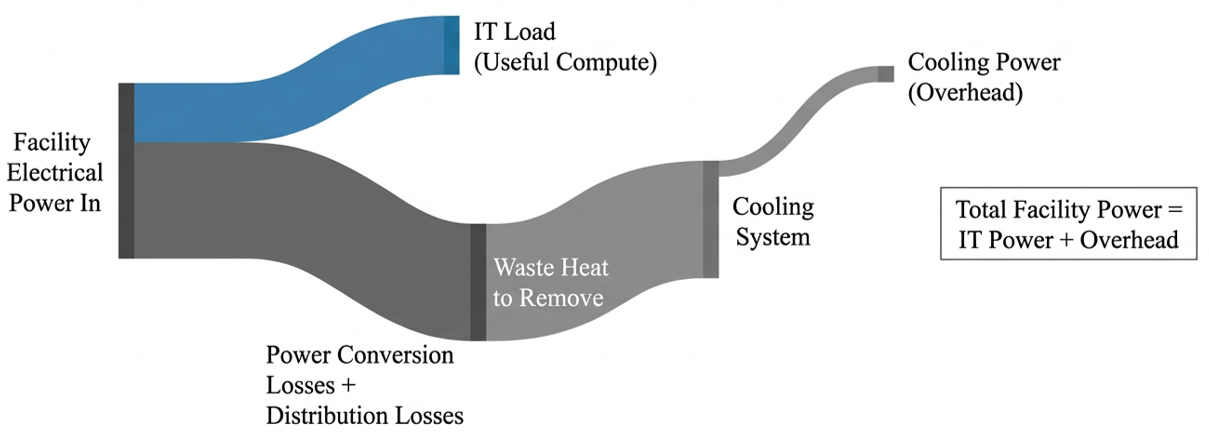}
\caption{Conceptual relationship between conversion losses, waste heat, and cooling overhead.}
\label{fig:fig2}
\end{figure}

\begin{figure*}[!t]
\centering
\includegraphics[width=\textwidth]{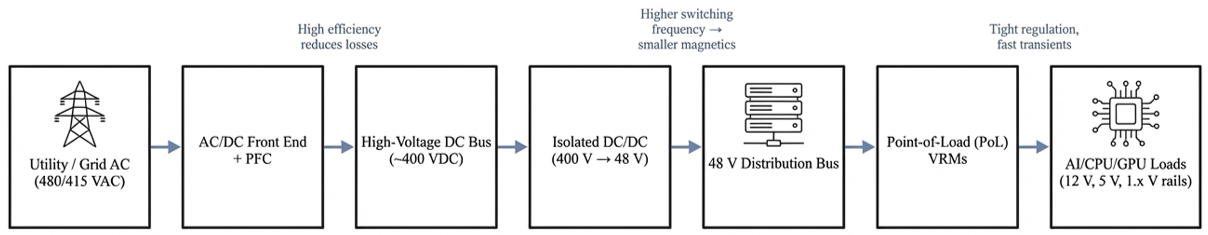}
\caption{Power conversion and distribution chain in an AI/data-center environment.}
\label{fig:fig1}
\end{figure*}

Artificial intelligence (AI) is rapidly reshaping the energy footprint of digital infrastructure \cite{sunkara2025power}. As training and inference workloads scale, data centers must deliver increasing electrical power to processors and accelerators while maintaining strict reliability, transient-response, and uptime requirements \cite{li2025ai}. This growth creates a critical power-electronics bottleneck upstream of the computing hardware: the efficiency, power density, and thermal behavior of the conversion and distribution chain that transfers energy from the utility interface to the processor point of load \cite{li2025ai}. As illustrated in Fig.~\ref{fig:fig1}, this chain commonly includes AC/DC rectification and power-factor correction, high-voltage bus regulation, isolated intermediate-bus conversion, and point-of-load regulation. Losses introduced at each stage appear primarily as waste heat, increasing both the input power required to support a given computational load and the cooling demand imposed on the facility \cite{zhang2022prediction}. The resulting interaction among conversion loss, waste-heat generation, and cooling overhead is conceptually summarized in Fig.~\ref{fig:fig2}.

In this context, advances in power electronics are not merely incremental component-level improvements; they can act as first-order levers on facility energy consumption and operational emissions \cite{li2025ai}. Higher conversion efficiency directly reduces electrical loss, while lower dissipated power reduces the thermal load that must be removed by cooling systems. Together, these effects can improve power usage effectiveness (PUE) and reduce the electricity required per unit of delivered computational work \cite{sunkara2025power}. For AI-oriented facilities operating continuously at high utilization, even sub-percentage-point improvements in stages that process the full rack or facility power can therefore translate into meaningful reductions in energy use, operating cost, and carbon intensity, particularly where the electricity supply remains partially fossil based.

\begin{figure*}[!t]
\centering
\includegraphics[width=\textwidth]{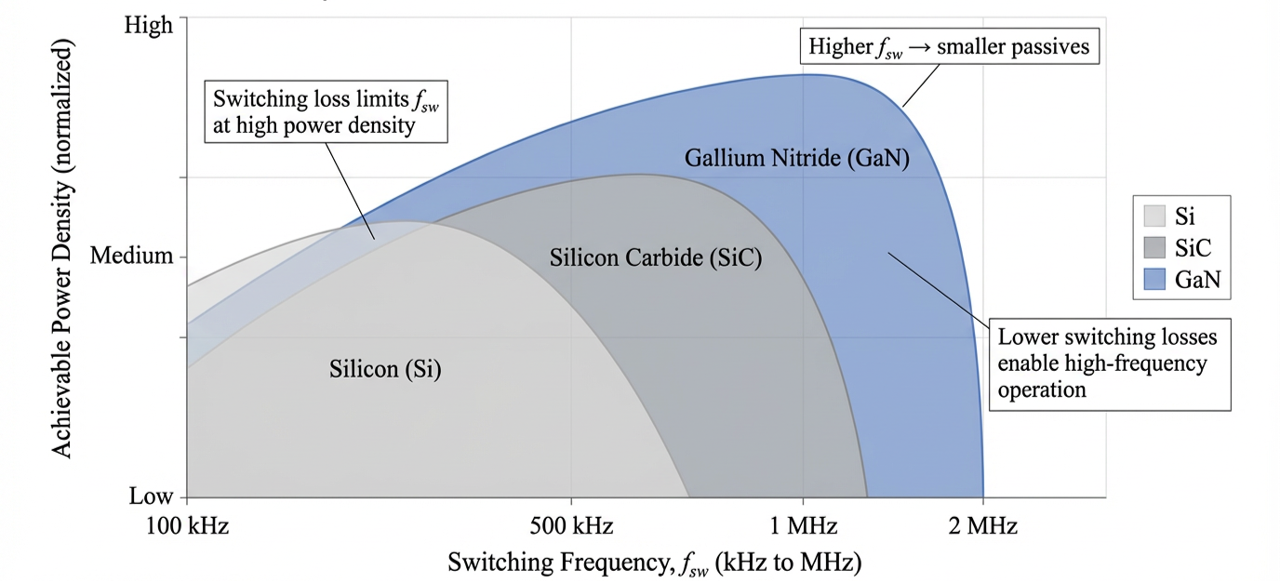}
\caption{Illustrative comparison of switching-frequency and power-density envelopes for Si, SiC, and GaN devices.}
\label{fig:fig3}
\end{figure*}

Silicon power devices have benefited from decades of technological optimization, but their performance becomes increasingly constrained as converter designs move toward higher switching frequency, higher volumetric power density, and tighter thermal margins \cite{chaudhary2023technology}. Increasing switching frequency can reduce the volume of magnetic and capacitive components, but it also raises switching loss, electromagnetic-interference (EMI) sensitivity, and stress associated with parasitic inductance and capacitance. These challenges are intensified by sustained 24/7 operation and the rapid load transients characteristic of processor and accelerator platforms. Consequently, wide-bandgap semiconductors are receiving increased attention because they can support higher electric fields and faster switching while maintaining competitive conduction loss and thermal stability \cite{chaudhary2023technology}. The qualitative relationship among switching frequency, achievable power density, and semiconductor technology is illustrated in Fig.~\ref{fig:fig3}.

Gallium nitride (GaN) has emerged as a particularly compelling wide-bandgap platform for next-generation power conversion \cite{musumeci2023gallium,chaudhary2023technology}. GaN has a bandgap of approximately $3.4\,\mathrm{eV}$, a critical breakdown field on the order of $3.0$--$3.5\,\mathrm{MV/cm}$, and high electron transport capability, including a higher saturated electron velocity than silicon. These properties enable devices that sustain high electric fields using compact active regions and that operate with reduced charge-related switching loss. At the converter level, GaN can reduce switching loss through low device capacitances, limited stored charge, and rapid voltage and current transitions, while low on-resistance designs can limit conduction loss. These attributes support increased switching frequency, reduced passive-component volume, and higher converter power density, consistent with the trend shown in Fig.~\ref{fig:fig3}.

The relevance of GaN to data-center power delivery is determined not only by its intrinsic material properties but also by the structural diversity of the available device technologies. GaN power devices span lateral and vertical conduction architectures, together with specialized structures designed to address threshold-voltage control, electric-field management, thermal resistance, bidirectional power flow, and system integration \cite{meneghini2021gan,yadlapalli2021advancements}. This diversity is important because the conversion stages in Fig.~\ref{fig:fig1} do not impose uniform requirements. Front-end power-factor-correction stages, isolated intermediate-bus converters, 48-V conversion stages, and processor-level regulators operate at different voltage classes, current levels, switching frequencies, conversion ratios, and thermal conditions. A portfolio of GaN architectures therefore provides a broader design space for matching device characteristics to stage-specific requirements while balancing efficiency, power density, controllability, and reliability \cite{meneghini2021gan,yadlapalli2021advancements}.

Lateral GaN/AlGaN high-electron-mobility transistors (HEMTs) are currently the most mature GaN devices for high-frequency, low-to-mid-voltage power conversion. Their operation is based on the polarization-induced two-dimensional electron gas (2DEG) formed at the AlGaN/GaN heterointerface, which provides a high-mobility conduction channel with low channel resistance \cite{meneghini2021gan}. In data-center converters, lateral GaN HEMTs are attractive because their low switching charge and absence of a conventional silicon MOSFET body diode can substantially reduce switching and reverse-recovery-related losses \cite{meneghini2017technology,yadlapalli2021advancements}. These characteristics are particularly valuable in high-frequency PFC, resonant DC/DC, intermediate-bus, and point-of-load stages, where converter efficiency and volumetric power density are both critical \cite{yadlapalli2021advancements}.

Vertical GaN devices provide a complementary pathway for voltage and current ranges in which lateral drift-region scaling becomes increasingly restrictive. By conducting current through the semiconductor bulk rather than parallel to the wafer surface, vertical architectures can scale blocking voltage through the drift-layer thickness while supporting higher current density and potentially more direct heat-extraction paths \cite{langpoklakpam2023vertical,gupta2022vertical}. These characteristics are relevant to higher-power stages such as power distribution units, uninterruptible power supplies, high-voltage DC/DC converters, and emerging high-voltage DC distribution architectures, where conduction loss, breakdown capability, and thermal stability must be maintained under sustained loading \cite{langpoklakpam2023vertical,gupta2022vertical}. The distinction between lateral and vertical GaN should therefore not be treated as a simple replacement hierarchy. Instead, the two architectures occupy complementary regions of the voltage--frequency--power design space and should be matched to the switching regime, thermal boundary conditions, and reliability requirements of the corresponding conversion stage \cite{meneghini2021gan,langpoklakpam2023vertical,gupta2022vertical} shown in Fig.~\ref{fig:fig1}.

Specialized GaN structures further broaden this design space. Trench-gate GaN MOSFET concepts can improve electric-field control and reduce specific on-resistance while supporting high breakdown voltage, providing a potential route toward compact, high-power-density conversion \cite{jaiswal2023optimized}. Bidirectional GaN HEMTs can reduce device count and conduction-path complexity in systems requiring bidirectional energy transfer, storage integration, AC switching, or advanced rack-level power management \cite{alharbi2021experimental}. However, device-level switching capability does not automatically translate into converter-level performance. Metallization, package architecture, gate-loop design, commutation-loop inductance, thermal interfaces, and EMI control strongly determine whether the intrinsic speed of GaN produces lower loss or instead produces excessive overshoot, ringing, and electrical stress. In high-density power modules, reduced parasitic inductance can limit switching overshoot and loss, while improved thermal paths can lower junction temperature and support reliable 24/7 operation \cite{chrzan2025gan}. These interactions directly affect the facility-level loss and cooling relationship illustrated in Fig.~\ref{fig:fig2}.

Manufacturability and scalability are similarly important for large-scale data-center deployment. GaN devices can be fabricated on silicon, silicon carbide, or native GaN substrates, with each platform providing different trade-offs among wafer cost, defect density, thermal performance, voltage capability, and manufacturing maturity \cite{yadlapalli2021advancements}. The ability to fabricate lateral GaN devices on large-diameter silicon substrates can improve manufacturing scalability and reduce barriers relative to approaches that rely exclusively on bulk GaN substrates. Nevertheless, total cost of ownership is determined not only by transistor price, but also by converter efficiency, package complexity, qualification requirements, cooling infrastructure, reliability, and replacement frequency \cite{yadlapalli2021advancements}.

Although GaN device physics, device reliability, converter implementation, and data-center energy efficiency have each received substantial attention, these topics are often treated separately. Device-oriented reviews generally emphasize material properties, gate structures, trapping, and breakdown behavior, whereas converter studies focus on individual topologies or demonstrations, and data-center studies typically evaluate facility energy use, PUE, and cooling. A stage-resolved synthesis is therefore needed to connect GaN device architecture and switching behavior to converter topology, operating domain, packaging requirements, and facility-level consequences.

Accordingly, this review examines how GaN material properties and architectural diversity can be translated into stage-specific benefits across the AI data-center power-delivery chain. The principal contributions are fourfold. First, the review develops a converter-oriented classification of lateral, vertical, and specialized GaN devices based on voltage scalability, switching behavior, conduction loss, thermal pathways, and reliability limitations. Second, it maps these device classes onto representative front-end PFC, isolated DC/DC, 48-V intermediate-bus, and point-of-load conversion stages. Third, it evaluates the package, gate-drive, EMI, thermal, manufacturing, and qualification constraints that determine whether device-level advantages are preserved at the converter and system levels. Fourth, it introduces a quantitative framework that connects cascaded conversion efficiency to electrical loss, waste-heat generation, cooling demand, and operational carbon impact. The objective is not to present GaN as a universal replacement for silicon or silicon carbide, but to identify the operating conditions under which device--topology--package co-optimization provides a defensible system-level advantage.

\begin{figure*}[!ht]
\centering
\includegraphics[width=.89\linewidth]{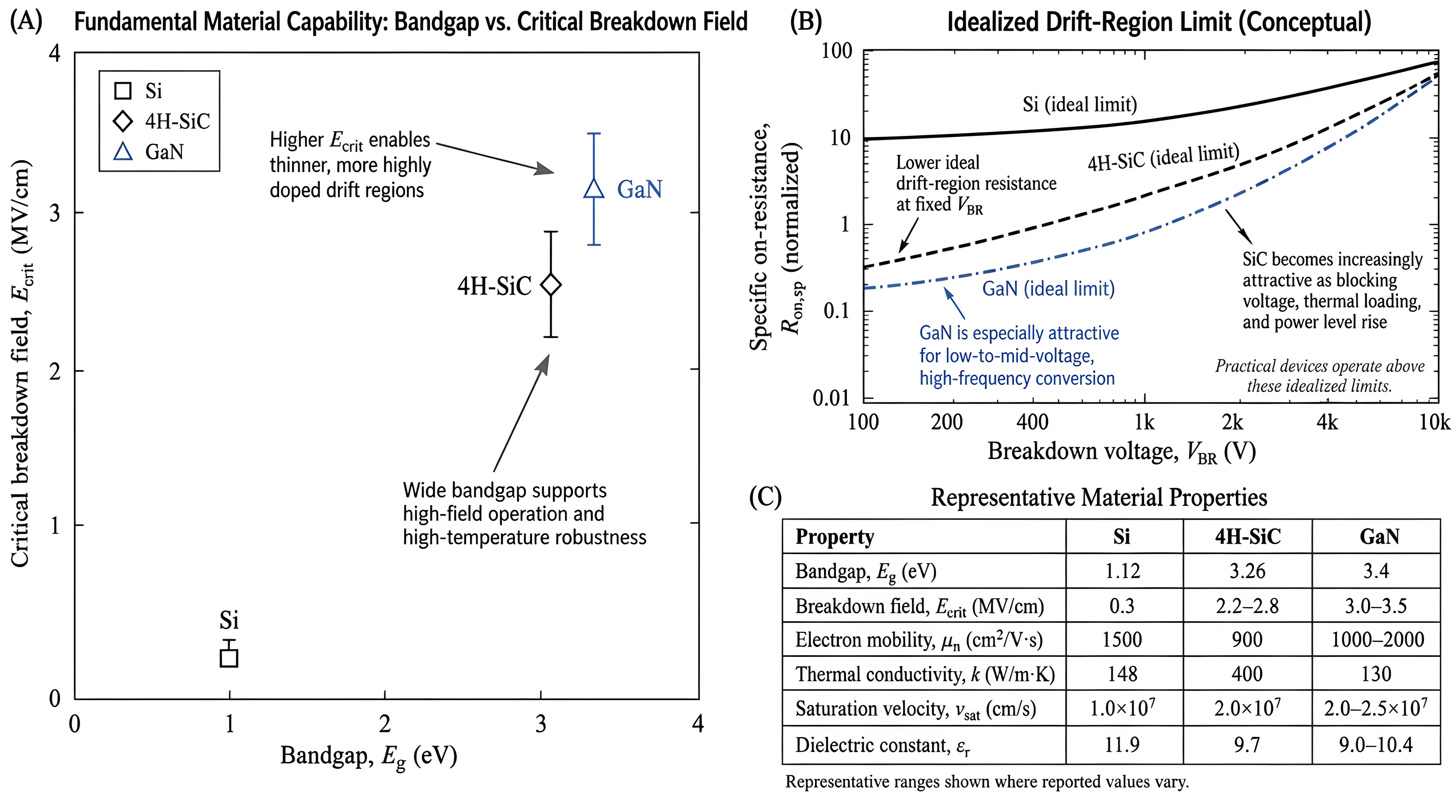}
\caption{Material performance comparison: (A) Fundamental relationship between breakdown field ($E_{crit}$) and bandgap ($E_g$); (B) Conceptual drift-region limits showing specific on-resistance ($R_{on,sp}$) versus breakdown voltage ($V_{BR}$); and (C) Summary of key material properties for Si, 4H-SiC, and GaN (data compiled from \cite{matsunami2020fundamental,ayalew2004sic,matocha2008challenges,zhang2025applicability,cittanti2022new,dai2026study}).}
\label{fig:material_comparison}
\end{figure*}

The material basis for the device-level comparisons used throughout the review is summarized in Fig.~\ref{fig:material_comparison}. The remainder of the paper progresses from material properties and device architectures to converter-stage deployment, manufacturing and reliability constraints, quantitative facility-level implications, and future research priorities for GaN-enabled AI power infrastructure.

\section{Converter-Relevant Material Background: GaN, Si, and SiC}
\label{sec:materials}

The performance potential of a power semiconductor is governed by how its intrinsic material properties translate into voltage-blocking capability, conduction loss, switching behavior, and thermal management. Compared with conventional silicon (Si), gallium nitride (GaN) supports higher electric fields, faster carrier transport, and reduced theoretical drift-region resistance, enabling high-efficiency and high-power-density conversion when these material advantages are preserved at the device, package, and converter levels \cite{rafin2023power}. As illustrated in Fig.~\ref{fig:material_comparison}(a), the wide bandgap and high critical electric field of GaN provide the physical basis for compact voltage-blocking regions \cite{yadlapalli2021advancements}. Together with its high carrier-transport capability, these properties make GaN particularly relevant to high-frequency power conversion in data-center and energy-system applications \cite{rafin2023power,yadlapalli2021advancements}.

\subsection{Breakdown Capability and Drift-Region Resistance}

As summarized in Fig.~\ref{fig:material_comparison}(c), GaN and 4H-SiC have substantially wider bandgaps and higher critical electric fields than Si \cite{rafin2023power,zhang2023status}. A higher critical field allows a power device to support a specified blocking voltage using a thinner and more heavily doped drift region. This reduces the theoretical specific on-resistance associated with voltage blocking, as illustrated conceptually in Fig.~\ref{fig:material_comparison}(b). GaN's critical electric field, which exceeds approximately $3~\mathrm{MV/cm}$, therefore provides a strong theoretical basis for compact devices with low drift-region resistance, while its high electron mobility and saturation velocity support rapid carrier transport \cite{islam2022reliability}.

The limits shown in Fig.~\ref{fig:material_comparison}(b) represent idealized unipolar drift-region behavior rather than the total on-resistance of a practical device. Actual resistance also includes contributions from the channel, access regions, contacts, substrate, current spreading, metallization, and package interconnects. In GaN HEMTs, trapping and self-heating can further increase the effective on-resistance under dynamic switching conditions. Consequently, the theoretical material limit should be interpreted as a comparison of intrinsic voltage-blocking potential rather than as a direct prediction of converter efficiency.

\subsection{Switching and Reverse-Conduction Behavior}

GaN's converter-level advantage is strongly associated with switching-related device metrics rather than with breakdown field alone. High carrier velocity, compact device geometry, and low charge storage can reduce gate-drive demand and voltage--current overlap during switching transitions. These characteristics can support higher switching frequency and smaller passive components, provided that increased magnetic loss, output-capacitance loss, package parasitics, and electromagnetic interference do not offset the semiconductor-level gain.

Lateral GaN high-electron-mobility transistors use a polarization-induced two-dimensional electron gas at the AlGaN/GaN heterointerface to obtain a high-mobility conduction channel with low channel resistance \cite{islam2022reliability,ma2019review}. Unlike a conventional Si MOSFET, a lateral GaN HEMT does not contain an intrinsic body diode with stored minority-carrier charge. It can therefore avoid conventional body-diode reverse-recovery loss, although reverse conduction still produces loss and the charging and discharging of output capacitance remain important during commutation. These characteristics are especially valuable in high-frequency PFC, resonant DC/DC, intermediate-bus, and point-of-load converters. GaN devices are also used in RF systems, although RF operation is outside the principal power-conversion scope of this review.

Commercial lateral GaN is particularly well established in approximately 100--650 V high-frequency conversion, with devices and converter demonstrations extending beyond this range. Its compatibility with silicon substrates can support large-diameter wafer processing and scalable manufacturing, although substrate choice, epitaxial buffer design, and defect management influence dynamic performance, thermal behavior, and reliability.

\subsection{Thermal Behavior and Practical Power Density}

Silicon carbide is particularly well suited to high-voltage, high-temperature, and high-power operating environments \cite{shi2023deep}. The thermal conductivity of 4H-SiC can approach $400~\mathrm{W/(m\cdot K)}$, providing a strong material-level heat-spreading advantage over both Si and GaN. This contributes to the use of SiC devices in converters that require high blocking voltage, substantial current capability, elevated junction temperature, and robust operation under severe electrical and thermal stress.

Bulk GaN has a lower thermal conductivity than 4H-SiC, as indicated in Fig.~\ref{fig:material_comparison}(c). Therefore, GaN's high achievable power density does not imply that heat removal is inherently easier. Faster switching and smaller die or package dimensions may reduce total loss while simultaneously increasing local heat flux. The realized junction temperature depends on the complete thermal path, including the epitaxial structure, substrate, die attach, package, printed-circuit board, heat spreader, and cooling system. Power density must therefore be evaluated together with thermal resistance and allowable junction temperature rather than inferred solely from semiconductor switching speed.

\subsection{Stage-Specific Positioning of GaN, SiC, and Si}

The relative suitability of Si, SiC, and GaN cannot be defined by a single voltage threshold. Silicon remains highly competitive in cost-sensitive and mature converter platforms, particularly where switching frequency and power density are moderate. Lateral GaN is especially attractive in low-to-mid-voltage stages where switching loss, passive-component size, transient response, and volumetric density dominate the design objective. Figure~\ref{fig:material_comparison}(b) therefore highlights a conceptual GaN advantage in mid-voltage operation, but this advantage depends on topology, switching mode, package parasitics, cooling, and operating frequency rather than on voltage alone.

SiC is commercially mature across multiple voltage classes, including 650-V devices as well as 1.2-kV and higher-voltage platforms. It is particularly advantageous when blocking-voltage margin, thermal conductivity, surge capability, high-temperature operation, and high-power robustness are more important than maximum switching frequency \cite{shi2023deep}. SiC MOSFETs and Schottky diodes are consequently widely used in electric-vehicle traction inverters, charging systems, renewable-energy converters, and grid-scale power electronics.

Emerging vertical GaN devices exploit the high critical field of GaN while conducting through the semiconductor thickness, providing a potential route toward higher blocking voltage, increased current capability, and improved voltage-area scaling \cite{langpoklakpam2023vertical}. Vertical GaN may therefore extend the useful range of GaN beyond the operating domains currently dominated by lateral HEMTs. However, its practical deployment remains constrained by native-substrate cost, epitaxial defect control, edge termination, processing maturity, packaging, and high-voltage qualification. Specialized architectures, including trench-gate GaN MOSFETs and bidirectional HEMTs, further tailor electric-field control, normally-off operation, and current-flow capability for specific converter requirements.

Taken together, GaN and SiC both offer substantial advantages over Si, but they occupy overlapping rather than strictly separated design spaces \cite{rafin2023power}. GaN is strongest where high-frequency switching, compact passive components, and high converter density are primary objectives, while SiC is particularly competitive where high voltage, high power, thermal robustness, and fault tolerance dominate. The preferred technology must therefore be selected using converter-level requirements, including voltage and current stress, switching mode, operating frequency, thermal boundary, reliability target, and cost. These material-level trade-offs provide the basis for the device-architecture discussion that follows, in which lateral, vertical, and specialized GaN structures are evaluated according to how effectively they convert intrinsic material capability into practical power-electronics performance.

\section{GaN Power-Device Architectures and Converter-Level Trade-Offs}
\label{sec:device_architecture_landscape}

\begin{figure*}[!t]
\centering
\includegraphics[width=.95\linewidth]{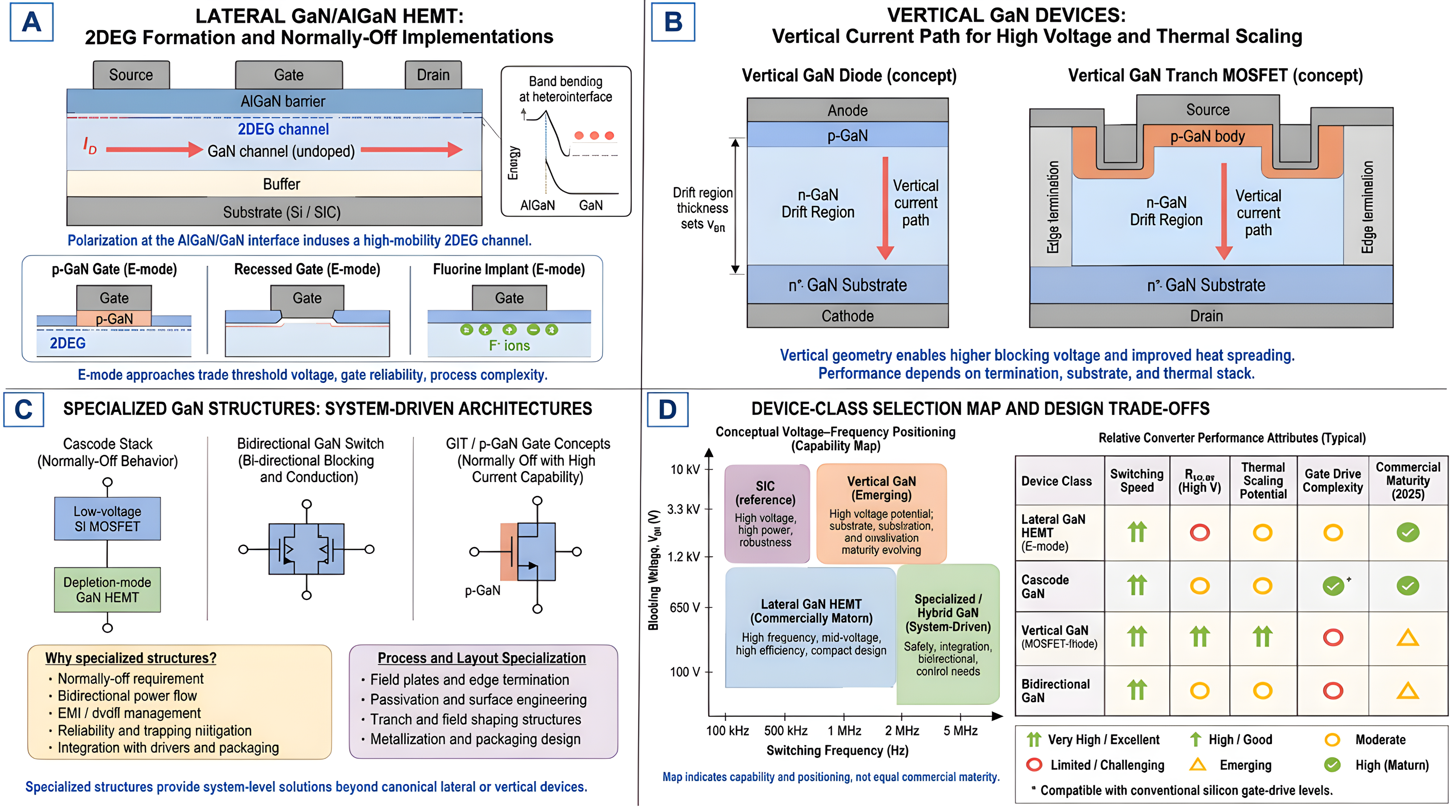}
\caption{Overview of GaN power-device architectures and their converter-level positioning.
(A) Lateral GaN/AlGaN HEMT structure and representative normally-off implementations, including p-GaN gate, recessed-gate, and fluorine-implant approaches, with polarization-induced 2DEG formation.
(B) Vertical GaN device concepts, including p--n diodes and trench-MOSFET structures with current flow through the bulk drift region.
(C) Specialized architectures, including cascode configurations, bidirectional switches, and GIT/p-GaN gate concepts.
(D) Conceptual device-class selection map illustrating voltage--frequency trade-offs, converter-level strengths, and architectural positioning. The map represents technological capability rather than equal commercial maturity among the device classes.}
\label{fig:GaN_architecture_landscape}
\end{figure*}

GaN power devices translate the intrinsic material properties discussed in Section~\ref{sec:materials} into practical voltage-blocking, conduction, switching, and thermal characteristics. Their importance to power electronics arises from the possibility of combining high conversion efficiency, elevated switching frequency, and high volumetric power density \cite{kinzer2022advancing}. These attributes are increasingly relevant as AI workloads and data-center electrification increase rack power, tighten thermal constraints, and place greater emphasis on converter efficiency and compact power delivery \cite{udabe2023gallium,boschee2024comments,shankar2024enhancing,sunkara2025power}. However, GaN's converter-level performance is determined not by material capability alone, but by the device architecture used to control the channel, sustain electric field, conduct reverse current, extract heat, and interface with the gate driver and package. Its high critical electric field and carrier-transport capability provide the basis for low-loss, high-frequency operation, particularly in low-to-mid-voltage conversion \cite{rafin2023power}.

A defining feature of the GaN platform is therefore its architectural diversity. Contemporary GaN power devices include: (i) lateral GaN/AlGaN HEMTs, which currently provide the most mature commercial route to high-frequency conversion; (ii) vertical GaN devices, which use bulk current flow to improve voltage-area scaling at higher blocking voltages; and (iii) specialized architectures that address normally-off operation, bidirectional conduction, field management, driver compatibility, and monolithic integration. Fig.~\ref{fig:GaN_architecture_landscape} summarizes these device classes and their principal converter-level trade-offs.

\subsection{Lateral GaN/AlGaN HEMTs and Normally-Off Implementations}
\label{subsec:lateral_hemt}

As illustrated in Fig.~\ref{fig:GaN_architecture_landscape}(A), lateral GaN/AlGaN HEMTs use polarization-induced charge at the AlGaN/GaN heterointerface to form a high-mobility two-dimensional electron gas (2DEG) \cite{nguyen2021piezotronic}. The 2DEG provides a low-resistance lateral conduction channel without intentional channel doping and supports rapid modulation of drain current during switching \cite{nguyen2021piezotronic}. The source, gate, and drain are arranged on the same surface, while the lateral access and drift regions determine the conduction path and contribute to the device voltage-blocking capability.

For practical power conversion, normally-off operation is required so that the device remains nonconducting at zero gate bias. Representative enhancement-mode implementations shown in Fig.~\ref{fig:GaN_architecture_landscape}(A) include p-GaN gate structures, recessed-gate geometries, and fluorine implantation beneath the gate \cite{ji2021ridge,chen2024research,kang2020charging}. Each approach modifies the electrostatics beneath the gate to deplete the 2DEG at zero bias and produce a positive threshold voltage. Their practical differences involve threshold-voltage magnitude and stability, gate leakage, allowable gate-voltage range, process complexity, and sensitivity to charge trapping.

Lateral GaN devices are commercially established in high-frequency power conversion because of mature GaN-on-Si and GaN-on-SiC processing, available enhancement-mode products, and favorable switching-related figures of merit. Their low switching charge and absence of a conventional minority-carrier body diode reduce commutation loss relative to many silicon implementations. However, reverse conduction still incurs a voltage-dependent loss, and output-capacitance charging and discharging remain important during hard switching. The realized converter advantage therefore depends on dead-time control, switching-node capacitance, package inductance, and the extent to which the topology provides soft switching.

As the required blocking voltage increases, the lateral drift region must generally be extended, increasing die area and exposing a greater surface region to high electric field. Surface and buffer trapping can produce dynamic on-resistance, current collapse, and time-dependent changes in conduction loss following high-voltage switching events \cite{kang2020charging}. Thus, lateral GaN is particularly attractive where switching-related loss and passive-component volume dominate, but its operation must be constrained by gate reliability, dynamic $R_{\mathrm{DS(on)}}$, electric-field management, and package-level parasitics.

\subsection{Vertical GaN Devices}
\label{subsec:vertical_gan}

Vertical GaN devices route current perpendicular to the wafer surface through a bulk drift region, as illustrated in Fig.~\ref{fig:GaN_architecture_landscape}(B) \cite{langpoklakpam2023vertical,he20221}. Unlike lateral HEMTs, whose blocking voltage is strongly influenced by lateral drift length and surface-field management, the blocking capability of a vertical device is governed primarily by drift-layer thickness, doping, junction design, and edge termination \cite{langpoklakpam2023vertical,duan20231}. This geometry provides a potential route to blocking voltages exceeding 1~kV while maintaining competitive specific on-resistance and limiting the increase in lateral die area \cite{langpoklakpam2023vertical,he20221,duan20231}.

Representative vertical structures include p--n diodes, current-aperture devices, trench-gate transistors, and vertical MOSFET concepts. The structures depicted in Fig.~\ref{fig:GaN_architecture_landscape}(B) illustrate how a bulk drift region and trench-based field control can combine vertical conduction with high-voltage blocking \cite{langpoklakpam2023vertical,kaminski2024vertical,duan20231}. Vertical current flow can reduce dependence on long surface drift regions, mitigate lateral current crowding, and distribute electric-field stress through the device thickness \cite{langpoklakpam2023vertical,kaminski2024vertical}.

The vertical geometry can also provide a more direct thermal path toward the substrate and die attach, but it does not automatically guarantee lower junction temperature. The realized thermal resistance depends on the conductivity and thickness of the substrate, active-region placement, backside contact, die attach, package architecture, and external cooling boundary. Vertical current alignment may improve current spreading and heat-flow geometry, but the complete thermal stack still determines performance under sustained power density \cite{langpoklakpam2023vertical,kaminski2024vertical}.

Vertical GaN remains less mature than commercial lateral GaN. Its adoption is constrained by native-substrate cost, limited wafer diameter, epitaxial defect density, p-type doping and activation, trench-interface quality, edge termination, and process reproducibility \cite{langpoklakpam2023vertical,kaminski2024vertical,ma2025low}. Native GaN substrates can reduce dislocation density and support high-field vertical structures, but their availability and cost currently limit high-volume deployment \cite{langpoklakpam2023vertical,kaminski2024vertical}. Vertical GaN should therefore be regarded as an emerging high-voltage platform rather than a direct near-term replacement for established lateral GaN or SiC technologies. Its strongest prospective role is in voltage and power ranges where lateral drift-region scaling becomes inefficient and where sufficient value exists to justify the additional substrate and qualification complexity \cite{langpoklakpam2023vertical,ma2025low}.

\subsection{Specialized, Cascode, and Bidirectional Architectures}
\label{subsec:specialized_structures}

Specialized GaN structures address converter requirements that cannot be represented solely by breakdown voltage or static on-resistance. These requirements include normally-off behavior, conventional gate-drive compatibility, bidirectional current blocking, stable threshold voltage, electric-field shaping, and functional integration \cite{udabe2023gallium}. Representative architectures shown in Fig.~\ref{fig:GaN_architecture_landscape}(C) include cascode configurations, bidirectional GaN switches, and gate-injection transistor or advanced p-GaN gate concepts \cite{islam2022reliability,wang2023review}.

A cascode combines a normally-on, high-voltage GaN HEMT with a low-voltage silicon MOSFET to produce composite normally-off behavior \cite{udabe2023gallium}. The low-voltage MOSFET controls the source potential of the depletion-mode GaN device, allowing the composite switch to be driven using gate-voltage levels familiar from silicon power converters. This can simplify adoption in existing converter platforms, but the two-device structure introduces additional capacitances, internal interconnect parasitics, reverse-conduction behavior, and transient interactions that must be considered during fast switching.

Bidirectional GaN switches are intended to control current and block voltage in both directions, making them relevant to AC switching, matrix converters, solid-state circuit breakers, energy-storage interfaces, and bidirectional power routing \cite{alharbi2021experimental}. A monolithic bidirectional structure can potentially replace two anti-series unidirectional transistors, reducing device count and current-path resistance. However, gate control, common-source behavior, protection, and independent voltage blocking in both polarities increase the complexity of the device and converter implementation.

Gate-injection transistors use hole injection from a p-type gate region to achieve normally-off operation and high current capability \cite{wang2023review}. Advanced p-GaN and gate-stack concepts similarly seek to improve threshold control, gate-current behavior, and conduction performance. Their practical suitability depends on maintaining a stable threshold voltage and gate characteristic throughout repetitive switching, high temperature, and long-duration bias stress.

Specialization also occurs through process and layout engineering. Metallization design, surface passivation, buffer engineering, field plates, edge termination, and trench geometries can shape the electric field, suppress trapping, and improve breakdown and reliability margins \cite{kozak2023stability,islam2022reliability}. These features can improve one performance dimension while adding capacitance, process steps, or field concentration elsewhere. Device optimization must therefore account for the intended switching topology and mission profile rather than treating breakdown voltage, static $R_{\mathrm{DS(on)}}$, or threshold voltage as isolated objectives \cite{kozak2023stability}.

\subsection{Architecture Selection and Technology Maturity}
\label{subsec:comparative_advantages_challenges}

The conceptual voltage--frequency positioning of the main GaN device classes is summarized in Fig.~\ref{fig:GaN_architecture_landscape}(D). Lateral GaN HEMTs are strongest in high-frequency, low-to-mid-voltage conversion, where switching charge, output capacitance, and compact passive components strongly influence converter performance \cite{udabe2023gallium}. Vertical GaN becomes increasingly attractive as voltage, current, and die-area scaling place greater emphasis on drift-region resistance, bulk field management, and high-power thermal pathways \cite{udabe2023gallium}. Cascode, bidirectional, and other specialized devices occupy system-driven positions in which normally-off behavior, conventional drive compatibility, bidirectional power flow, integration, or protection functionality is decisive \cite{udabe2023gallium}.

From a converter-loss perspective, lateral GaN is advantageous when transition loss, gate charge, commutation behavior, and device capacitances represent a significant fraction of total loss. Vertical GaN has the potential to become advantageous when high-voltage drift-region resistance, current spreading, and die-area scaling become limiting factors \cite{cittanti2022new}. Specialized structures do not necessarily provide the lowest intrinsic loss; instead, they may offer system-level value by reducing device count, simplifying the driver, enabling bidirectional operation, improving normally-off control, or integrating functions that would otherwise require discrete components \cite{udabe2023gallium}.

These classes also differ substantially in readiness. Enhancement-mode lateral GaN HEMTs are commercially mature and already used in PFC, isolated DC/DC, intermediate-bus, and compact power-supply applications. Cascode GaN is also commercially available, although its hybrid structure produces different switching and reverse-conduction behavior from a single-chip enhancement-mode HEMT. Monolithic bidirectional devices, advanced power integrated circuits, and vertical GaN transistors occupy emerging or early-commercialization stages, depending on the specific structure and voltage class.

For continuously operated AI data centers, device selection must therefore balance efficiency, switching frequency, voltage capability, package parasitics, thermal resistance, fault behavior, manufacturability, and long-term stability \cite{musumeci2023gallium}. A heterogeneous device landscape is the most credible deployment pathway: commercially mature lateral GaN for prevalent high-frequency 400--650-V stages, specialized or hybrid devices where control and integration requirements dominate, and vertical GaN as a longer-term option for higher-voltage and higher-power conversion after manufacturing and qualification barriers are resolved \cite{musumeci2023gallium}. This architecture-based framework provides the basis for the stage-specific data-center power-chain mapping developed in the following section.

\section{System-Level Mapping of GaN Devices to Data-Center Power Conversion}

The device-level advantages of GaN become most consequential when evaluated across the complete data-center power-delivery chain, in which multiple conversion stages are cascaded across distinct voltage domains and their losses combine to determine the end-to-end efficiency \cite{rahman2023review}. GaN's wide bandgap and high critical electric field reduce the drift-region penalty associated with voltage blocking, while its high electron transport capability supports rapid switching with reduced charge-related loss \cite{meneghini2021gan}. However, favorable material properties or device figures of merit do not independently guarantee converter-level improvement. System benefit is realized only when the GaN architecture, voltage rating, package, gate drive, commutation loop, switching strategy, and thermal path are matched to the operating requirements of a specific conversion stage. These requirements must also be satisfied under the continuous loading, fast transients, and high availability expected in 24/7 data-center operation \cite{chrzan2025gan}. Figure~\ref{fig:dc_chain_map} maps representative GaN device classes onto the data-center power chain, while Table~\ref{tab:dc_chain_map} summarizes representative operating domains, reported performance, dominant design constraints, and technology maturity.

\begin{table*}[!ht]
\centering
\scriptsize
\setlength{\tabcolsep}{3.2pt}
\renewcommand{\arraystretch}{1.18}
\caption{Representative operating domains and demonstrated performance metrics across the data-center power chain, with implications for GaN device-class selection. Values are representative examples from recent reference designs and standards (data compiled from \cite{OpenComputeORv3,TEORv3Power,yu2024novel,InfineonGSEVB,EPCbp092025,ahmed2019gan,baek2020lego,shehabi20242024,UptimeInstitute2024,ASHRAETC92016}).}
\label{tab:dc_chain_map}
\begin{tabular}{
p{2.15cm}
p{2.15cm}
p{1.65cm}
p{2.05cm}
p{3.15cm}
p{2.65cm}
p{2.40cm}}
\toprule
\textbf{Power-chain stage} &
\textbf{Voltage domain} &
\textbf{Power scale} &
\textbf{Switching regime} &
\textbf{Representative performance and status} &
\textbf{Dominant converter-level constraints} &
\textbf{GaN implication} \\
\midrule

Rack distribution
(ORV3 48-V bus) &
46--52 VDC distribution &
Rack- and shelf-level current distribution &
N/A; passive distribution &
OCP ORV3 specifies a 46.0--52.0 VDC IT-gear interface rated at 100 A continuous. Ecosystem documentation describes distribution capability approaching $\sim$1000 A at the output-connector level. \textit{Status: deployed standard and ecosystem architecture.} &
Busbar and connector resistance, current sharing, protection coordination, copper loss, and conversion placement &
The 48-V bus reduces distribution $I^{2}R$ loss relative to lower-voltage buses and moves high-ratio conversion closer to the load, creating demand for compact 48-V-to-PoL converters. \\

AC/DC front end and PFC &
90--264 VAC $\rightarrow$ approximately 400 VDC &
3--3.2 kW module class &
Approximately 65 kHz to 500 kHz in representative designs &
A 3-kW bridgeless totem-pole PFC reference design reports 99\% peak efficiency with a listed maximum switching frequency of 65 kHz. A 3.2-kW interleaved critical-conduction-mode GaN totem-pole design reports 500-kHz operation and 99.3\% peak efficiency. \textit{Status: reference designs and demonstrated prototypes.} &
Hard commutation, $C_{\mathrm{oss}}$ energy, reverse conduction, common-mode EMI, current sensing, gate-loop inductance, and line-cycle thermal variation &
Lateral GaN is strongly positioned because fast switching and the absence of conventional body-diode reverse recovery support high-efficiency totem-pole operation. Package parasitics, EMI, and gate-voltage margin remain decisive. \\

Isolated DC/DC conversion after PFC &
400 VDC $\rightarrow$ 50 VDC
(48-V-class bus) &
Up to 5.5 kW demonstrated &
Resonant operation near 1 MHz &
An input-series-output-parallel LLC converter has been demonstrated at up to 5.5 kW between a 400-VDC input and a 50-V bus, with a resonant frequency near 1 MHz, 98.5\% peak efficiency, and approximately 97.5--98\% full-load efficiency depending on the implementation. \textit{Status: laboratory prototype.} &
Circulating current, transformer and resonant-tank loss, dead-time control, soft-switching range, isolation capacitance, package parasitics, and heat extraction &
High-frequency lateral GaN supports compact magnetics and modular isolation, but the system benefit depends on maintaining soft switching and controlling magnetic, capacitive, and interconnect losses. \\

48-V VRM first stage
(fixed-ratio DCX) &
48 V $\rightarrow$ intermediate bus using fixed ratios such as 4:1 or 8:1 &
Approximately 900 W continuous &
Approximately 1 MHz and above &
An unregulated LLC-DCX with integrated magnetics has demonstrated 4:1 and 8:1 conversion ratios at 900 W continuous output, with maximum efficiencies of 98.4\% and 98.0\%, respectively. \textit{Status: demonstrated converter prototype.} &
Transformer integration, winding and termination loss, current sharing, resonant-tank tolerance, board-level thermal spreading, and package footprint &
Lateral GaN enables MHz-class switching and magnetic integration, supporting conversion on the board or near accelerator packages where footprint and transient performance are critical. \\

48 V $\rightarrow$ processor-core PoL regulation
(hybrid or multiphase) &
48 V $\rightarrow$ approximately 1.5 V at very high current &
300 A demonstrated &
Switched-capacitor stage near 100 kHz and multiphase buck stage near 1 MHz &
The LEGO-PoL architecture combines a switched-capacitor stage operating near 100 kHz with a multiphase buck stage near 1 MHz. A representative implementation uses three stacked 16-V, 100-A submodules to provide 48-V-to-1.5-V, 300-A conversion. \textit{Status: research prototype.} &
Extreme conversion ratio, conduction loss, capacitor charge-transfer loss, multiphase current sharing, fast load transients, interconnect resistance, and local heat flux &
Specialized and hybrid architectures divide the conversion ratio across stages. GaN is most valuable where MHz switching, low charge, compact integration, and rapid transient response outweigh the associated gate-drive and layout complexity. \\

Emerging HVDC distribution for AI racks &
$\pm$400 V or 800-V-class DC distribution $\rightarrow$ local conversion &
Facility- and rack-level architectural trend &
Topology dependent; includes HVDC bus conversion, isolation, intermediate-bus conversion, and PoL regulation &
AI power-architecture roadmaps include $\pm$400-V and 800-V-class DC buses. Modular isolated stages can be extended to these buses through additional series-connected input modules. \textit{Status: emerging or prospective architecture rather than widespread deployment.} &
DC fault interruption, insulation coordination, protection selectivity, isolation, high-voltage packaging, qualification, and converter modularity &
Higher-voltage distribution may increase the long-term relevance of vertical GaN, but near-term adoption depends on device maturity, edge termination, package isolation, fault robustness, and system-level qualification. \\

\bottomrule
\end{tabular}
\end{table*}

\subsection{Stage Requirements and Device-Class Selection}

\begin{figure*}[t]
\centering
\includegraphics[width=\textwidth]{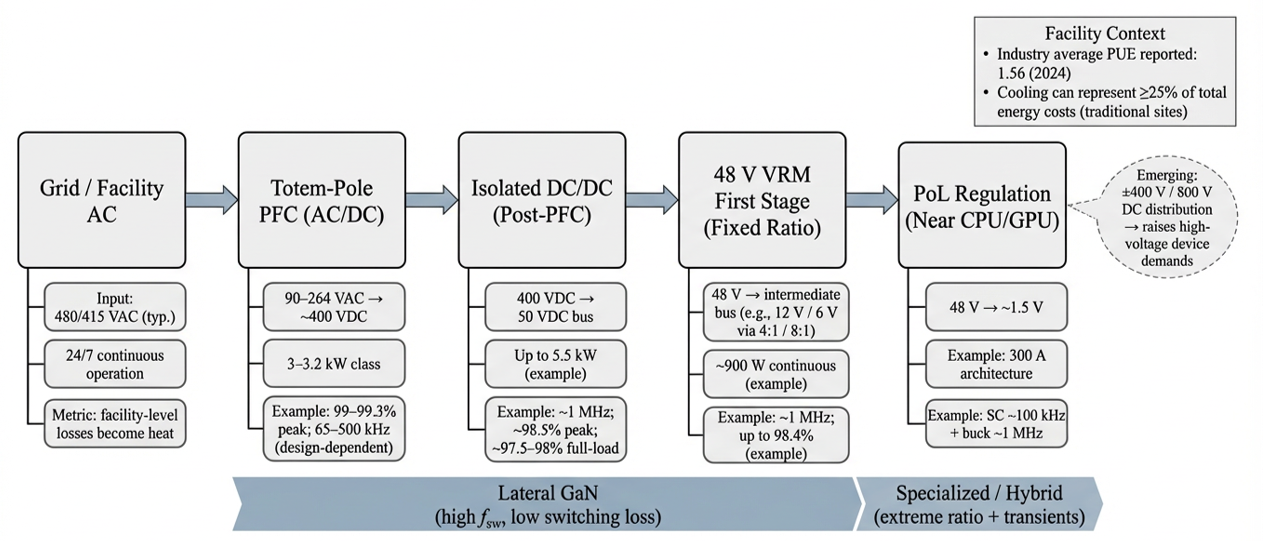}
\caption{Mapping of GaN device classes onto a representative data-center power-delivery chain using stage-level operating domains and reported performance metrics. Each block identifies the voltage domain, power scale, switching regime, and a representative efficiency from the facility AC input to point-of-load regulation. The lower band identifies stages in which lateral GaN is favored for high-frequency, low-loss conversion and stages in which specialized or hybrid architectures address extreme conversion ratios and fast transient requirements near CPUs and GPUs. The inset summarizes the facility context, including PUE, cooling demand, and emerging high-voltage DC distribution that may increase future high-voltage device requirements.}
\label{fig:dc_chain_map}
\end{figure*}

A data center converts incoming AC power into one or more regulated DC buses and ultimately into tightly controlled point-of-load rails for processors, memory, storage, and networking hardware \cite{lee2022design,ma2019review}. The conversion stages differ substantially in voltage, current, power throughput, switching frequency, isolation requirement, conversion ratio, transient response, and allowable thermal impedance. Consequently, device selection cannot be based on a single semiconductor figure of merit. Figure~\ref{fig:dc_chain_map} provides an at-a-glance mapping of representative stages and operating regimes, while Table~\ref{tab:dc_chain_map} summarizes the corresponding voltage domains, power scales, switching regimes, reported efficiencies, and implementation maturity \cite{lee2022design,ma2019review}.

For a representative stage, the total converter loss may be expressed conceptually as

\begin{equation}
P_{\mathrm{loss,stage}}
=
P_{\mathrm{semi}}
+
P_{\mathrm{mag}}
+
P_{\mathrm{cap}}
+
P_{\mathrm{int}}
+
P_{\mathrm{aux}},
\label{eq:stage_loss}
\end{equation}

where $P_{\mathrm{semi}}$ is semiconductor loss, $P_{\mathrm{mag}}$ is magnetic-component loss, $P_{\mathrm{cap}}$ is capacitor and charge-transfer loss, $P_{\mathrm{int}}$ is package, connector, busbar, and PCB interconnect loss, and $P_{\mathrm{aux}}$ includes gate-drive, control, sensing, and auxiliary-supply consumption. The semiconductor contribution can be further separated as

\begin{equation}
P_{\mathrm{semi}}
=
P_{\mathrm{cond}}
+
P_{\mathrm{tr}}
+
P_{C_{\mathrm{oss}}}
+
P_{\mathrm{gate}}
+
P_{\mathrm{rev}},
\label{eq:semiconductor_loss}
\end{equation}

where $P_{\mathrm{cond}}$ represents channel and reverse-conduction loss, $P_{\mathrm{tr}}$ represents voltage-current overlap during switching transitions, $P_{C_{\mathrm{oss}}}$ represents output-capacitance charging and discharging loss not recovered by the topology, $P_{\mathrm{gate}}$ represents gate-drive loss, and $P_{\mathrm{rev}}$ represents loss associated with reverse conduction and commutation. The relative importance of these terms changes with topology and operating point. Therefore, increasing switching frequency does not automatically improve efficiency: reductions in passive-component volume can be offset by increased semiconductor switching loss, magnetic core and winding loss, common-mode current, EMI-filter requirements, and local heat flux.

Lateral GaN/AlGaN HEMTs are best aligned with low-to-mid-voltage stages in which switching charge, transition loss, and commutation behavior strongly influence total loss. Their polarization-induced two-dimensional electron gas supports high mobility and low channel resistance while enabling rapid switching \cite{yadlapalli2021advancements,keshmiri2020current}. These characteristics make lateral devices attractive for PFC, isolated DC/DC, intermediate-bus, and near-load conversion operating from tens of kilohertz into the megahertz regime \cite{lee2022design,ma2019review,frivaldsky2020evaluation,faizan2023design}. Within this class, the choice among enhancement-mode, depletion-mode, and cascode implementations affects fail-safe behavior, gate-drive compatibility, reverse conduction, parasitic interaction, and response under abnormal conditions \cite{yadlapalli2021advancements,buffolo2024review}.

Vertical GaN devices become increasingly relevant when breakdown-voltage scaling, current density, and conduction loss make lateral drift-region extension inefficient \cite{langpoklakpam2023vertical,buffolo2024review}. Because current flows through the semiconductor thickness, the voltage-blocking region can be scaled predominantly through epitaxial thickness rather than lateral die dimension, offering a potential route to higher blocking voltage with reduced area penalty \cite{langpoklakpam2023vertical}. Vertical current flow may also provide a more direct path toward the substrate, although the realized thermal advantage depends on the substrate, die attach, package, and cooling boundary. As higher-voltage DC distribution architectures develop, vertical GaN may become relevant to high-voltage bus conversion and protection. However, this remains a prospective application subject to high-voltage edge termination, defect control, native-substrate cost, package insulation, surge capability, and mission-profile qualification \cite{langpoklakpam2023vertical,buffolo2024review}.

Specialized GaN devices address system constraints that cannot be resolved through voltage rating or on-resistance alone. Trench structures and field-management features can reduce electric-field crowding in high-voltage devices \cite{buffolo2024review}. Bidirectional GaN devices can provide controlled current flow in both directions and may reduce the number of series devices required in AC switching, storage interfaces, and bidirectional bus converters \cite{frivaldsky2020evaluation}. Advanced gate structures target positive and stable threshold voltage, low leakage, and improved gate reliability during sustained operation \cite{yadlapalli2021advancements,buffolo2024review}. Such architectures are most valuable when they resolve a specific converter bottleneck, including bidirectional operation, extreme conversion ratio, normally-off safety, switching-node integration, or fast transient response \cite{frivaldsky2020evaluation,faizan2023design,buffolo2024review}.

\subsection{Front-End AC/DC Conversion and Power-Factor Correction}

The front-end AC/DC stage processes essentially the full electrical power delivered to the downstream rack or power shelf. Consequently, even a small reduction in its loss can produce a meaningful absolute decrease in converter heat generation. Totem-pole and bridgeless PFC topologies are particularly relevant because they reduce the number of semiconductor drops in the primary current path. However, their high-frequency switching legs impose demanding commutation conditions, particularly near portions of the AC line cycle where current is low or changes direction.

Lateral GaN is well suited to these stages because it avoids the conventional body-diode reverse-recovery behavior of silicon superjunction MOSFETs and can reduce transition loss through low switching charge. The resulting benefit is not unconditional. Output-capacitance energy, reverse-conduction voltage drop during dead time, common-source inductance, gate-loop ringing, and common-mode EMI can become dominant as the switching edges are accelerated. The representative PFC systems in Table~\ref{tab:dc_chain_map} therefore illustrate two different design directions: moderate-frequency operation optimized for high efficiency and lower EMI burden, and substantially higher-frequency operation intended to increase power density and reduce passive-component size. In both cases, the useful GaN operating point is determined by a converter-level optimum rather than by the maximum switching speed of the transistor.

\subsection{Isolated High-Voltage-to-48-V Conversion}

The isolated DC/DC stage converts the regulated high-voltage bus into a 48-V-class distribution rail and must simultaneously satisfy efficiency, galvanic isolation, transformer utilization, thermal density, and hold-up or ride-through requirements. Resonant topologies such as LLC converters are attractive because zero-voltage switching can reduce the turn-on loss associated with transistor output capacitance. GaN enables operation at hundreds of kilohertz or into the megahertz range, allowing reductions in transformer and resonant-component volume.

At these frequencies, the semiconductor may no longer be the only or even the dominant source of loss. Transformer core loss, proximity and skin effects in windings, termination resistance, circulating resonant current, isolation capacitance, synchronous-rectifier timing, and PCB interconnect loss become increasingly important. The 5.5-kW example in Table~\ref{tab:dc_chain_map} demonstrates the feasibility of megahertz-class isolated conversion, but it also illustrates why high switching frequency must be accompanied by integrated magnetics, low-inductance packaging, controlled resonant operation, and an effective thermal path. The value of GaN in this stage is therefore its ability to expand the feasible design space for soft-switched, high-density conversion rather than to eliminate the underlying magnetic and thermal constraints.

\subsection{48-V Intermediate-Bus and Point-of-Load Conversion}

The final stages of the data-center power chain must convert a 48-V distribution bus into processor rails that can be near or below 1 V while supplying currents of hundreds of amperes. This extreme conversion ratio cannot generally be addressed efficiently by increasing switching frequency in a conventional single-stage buck converter. Instead, fixed-ratio DC transformers, switched-capacitor stages, multiphase buck converters, and hybrid architectures divide the voltage transformation and regulation functions across multiple sub-stages.

GaN can improve these converters through low switching charge, low package inductance, high-frequency capability, and the potential for close integration with magnetic and capacitive components. However, near-load conversion is frequently dominated by conduction and interconnect loss because the output current is extremely high. Package resistance, PCB copper, vias, connectors, inductors, current-sharing imbalance, and the physical distance between the regulator and accelerator package can therefore limit the total benefit. The DCX and LEGO-PoL examples in Table~\ref{tab:dc_chain_map} show how fixed-ratio and hybrid approaches can reduce the burden placed on the final regulating stage. In these architectures, GaN is most valuable when it enables a reduction in conversion-stage count, passive volume, or transient-response penalty without introducing excessive charge-transfer, gate-drive, or thermal loss.

\subsection{Emerging High-Voltage DC Distribution}

Higher-voltage DC distribution is being considered as a means to reduce current and conductor loss as rack power increases. Architectures based on $\pm$400 V or 800-V-class buses can reduce distribution current for a given power level but transfer greater responsibility to insulation coordination, DC fault interruption, isolation, protection selectivity, and local high-ratio conversion. These systems should be distinguished from currently deployed 48-V rack architectures because their protection and qualification ecosystems remain under development.

Lateral GaN may remain applicable in modular or series-connected converter cells, while vertical GaN could eventually provide a more direct device pathway for high-voltage conversion. Nevertheless, the adoption of vertical GaN in this domain is not determined by breakdown voltage alone. High-voltage package insulation, edge termination, surge robustness, short-circuit behavior, thermal extraction, defect density, manufacturing yield, and qualification maturity must all be addressed before device-level capability can translate into mission-critical deployment.

\subsection{Efficiency, Waste Heat, Cooling Load, and PUE}

Within the carbon-neutral data-center framework, converter efficiency is relevant not only as an electrical metric but also as a driver of heat generation and facility overhead \cite{chaudhary2023technology,yeboah2025wide}. Higher conversion efficiency reduces the electrical power dissipated in semiconductor devices, magnetics, capacitors, and interconnects. A portion of this reduction can also decrease the cooling power required to maintain acceptable component and air temperatures. The magnitude of the facility-level benefit depends on the stage power throughput, load profile, cooling-system coefficient of performance, climate, airflow architecture, and interaction with other facility loads.

GaN can reduce switching and conduction loss when it is deployed within an appropriate voltage, frequency, and topology domain \cite{keshmiri2020current,zhang2022optimal}. In lateral GaN converters, low switching charge and rapid transitions can improve stage efficiency and enable smaller passive components. These benefits are preserved only when the package, gate driver, layout, magnetic design, and control strategy prevent parasitic loss and excessive EMI. Under these conditions, lower semiconductor and converter loss can reduce local component temperature and thermal-management demand \cite{han2025thermal}, particularly in front-end PFC and isolated conversion stages that process large portions of the rack power.

The efficiency of a cascaded power chain is multiplicative. Improving any individual stage increases the end-to-end chain efficiency, but the physical propagation of the benefit depends on the stage location. An improvement in a downstream stage reduces the power demanded from every upstream stage for a fixed delivered load, whereas an improvement in an upstream stage primarily reduces the input power and heat generated locally at that stage. Stages that process the full rack power generally provide greater absolute leverage than converters serving only a small load branch. Thus, the benefit should be evaluated from the complete chain rather than inferred from peak device or single-stage efficiency alone \cite{chaudhary2023technology}.

The cooling consequence is similarly site dependent. A reduction in converter heat does not produce a universal or fixed reduction in PUE because PUE includes cooling, power distribution, lighting, controls, and other facility overheads. GaN should therefore be positioned as an enabling technology that can reduce conversion loss and local thermal density, supporting improved facility efficiency under appropriate operating conditions \cite{han2025thermal}, rather than as a device substitution that guarantees a predetermined reduction in PUE or carbon emissions. The quantitative relationship among cascaded efficiency, electrical loss, cooling demand, and carbon intensity is developed separately in the facility-level analysis.

\subsection{Practical Adoption Constraints in 24/7 Data-Center Operation}

Despite its favorable device characteristics, GaN deployment in mission-critical infrastructure remains constrained by gate-drive compatibility, $\mathrm{d}v/\mathrm{d}t$ management, EMI, package parasitics, protection behavior, and long-term qualification under continuous operation \cite{kozak2023stability}. These factors should be evaluated using application-relevant electrical and thermal mission profiles rather than only static device ratings.

Gate-drive margin is a central concern. Enhancement-mode GaN provides normally-off behavior at zero gate bias, but its relatively narrow allowable gate-voltage range and sensitivity to loop inductance require careful selection of the driver, turn-on and turn-off impedances, negative-bias strategy, and local decoupling \cite{heumesser2023cascode,bau2020cmos}. Cascode GaN can improve compatibility with conventional gate-drive levels, but the dynamic interaction between the low-voltage silicon MOSFET and depletion-mode GaN device must be validated during switching transitions, reverse conduction, startup, and fault events. High $\mathrm{d}v/\mathrm{d}t$ can produce Miller-induced false turn-on, gate ringing, common-mode current, and drain-voltage overshoot when gate-loop and commutation-loop inductances are not tightly controlled. Low-inductance packaging and co-located gate drivers are therefore central to realizing the high-frequency advantage of GaN \cite{chrzan2025gan,bu2020gan}.

Packaging is not a secondary implementation detail; it determines whether the intrinsic device capability is preserved in the converter \cite{chrzan2025gan}. Parasitic inductance influences voltage overshoot, current ringing, switching loss, and EMI, while parasitic capacitance controls common-mode displacement current and may increase isolation or filter requirements. At the same time, the package must provide a sufficiently low thermal impedance while maintaining dielectric isolation, mechanical integrity, manufacturability, and compatibility with high-volume assembly. These requirements are especially stringent in high-density power shelves and near-processor converters, where airflow can be restricted and local heat flux is high.

Reliability under continuous load remains a primary adoption criterion. Dynamic on-resistance and charge trapping can increase conduction loss after high-voltage switching events, while threshold-voltage drift, gate degradation, and package wear-out can alter converter behavior over time \cite{kozak2023stability}. Thermal cycling can fatigue die-attach layers, metallization, solder joints, and substrate interfaces even when the average operating temperature remains within specification. Qualification for data-center use should therefore combine conventional static stress testing with system-representative evaluation of repetitive switching, line and load transients, startup and shutdown, overload, thermal cycling, current sharing, EMI compliance, and fault recovery. The required test envelope should be derived from the voltage, current, switching, and thermal domains summarized in Table~\ref{tab:dc_chain_map} and Figure~\ref{fig:dc_chain_map}, rather than from a single nominal operating point.

\section{Manufacturing, Packaging, Reliability, and Supply-Chain Constraints}
\label{sec:manufacturing_constraints}

GaN deployment in energy- and carbon-conscious data centers ultimately depends on whether device-level performance can be reproduced at high manufacturing volume, integrated into low-parasitic and thermally robust packages, and qualified for continuous mission-critical operation. Substrate availability, epitaxial yield, metallization compatibility, package assembly, and supplier concentration therefore influence not only device cost but also converter efficiency, reliability, replacement frequency, and total cost of ownership \cite{yeboah2025wide,lu2025design}. Table~\ref{tab:gan_mfg_supply_slim} summarizes representative quantitative anchors related to substrate scalability, packaging, qualification, yield, and embodied impact.

\begin{table*}[!t]
\centering
\caption{Representative manufacturing, packaging, reliability, and supply-chain considerations relevant to GaN deployment in data-center power conversion. Quantitative values are screening-level anchors from publicly available technical references, datasheets, market reports, and reliability documentation rather than universal values for all GaN processes or products.}
\label{tab:gan_mfg_supply_slim}

\scriptsize
\setlength{\tabcolsep}{3pt}
\renewcommand{\arraystretch}{1.16}

\begin{threeparttable}
\begin{tabularx}{\textwidth}{
>{\raggedright\arraybackslash}p{2.50cm}
>{\raggedright\arraybackslash}p{4.15cm}
>{\raggedright\arraybackslash}X
>{\raggedright\arraybackslash}p{1.75cm}}
\toprule
\textbf{Manufacturing lever} &
\textbf{Representative quantitative anchor} &
\textbf{Implication for converter deployment} &
\textbf{Ref.} \\
\midrule

\multicolumn{4}{l}{\textbf{Substrate platforms and upstream supply}} \\

GaN-on-Si &
Publicly described manufacturing platforms include 200-mm wafers, with 300-mm GaN-on-Si milestones also reported. &
Provides the strongest leverage from existing Si-fabrication infrastructure. Scaling challenges shift toward epitaxial stress, wafer bow, crack control, defect management, and process uniformity, all of which influence yield and dynamic reliability. &
\cite{universitywafer_silicon_wafers,universitywafer_gan_on_si_epitaxy,okmetic_rf_gan_substrate_wafers,navitas_200mm_gan_psmc_2025,xfab_gan_on_si_foundry_2025,infineon_gan_technology,infineon_300mm_gan_press_2024,gansystems_gs66508t_datasheet_2020} \\

GaN-on-SiC &
SiC substrates are commercially offered at 150 and 200~mm. One supplier reports thermal conductivity near 370~W/(m$\cdot$K). &
Provides greater substrate-level thermal headroom and a favorable lattice relationship, but increases exposure to SiC wafer cost, capacity constraints, and competing demand from automotive and grid-power markets. &
\cite{coherent_sic_materials_datasheet,trendforce_sic_price_war_2025,navitas_sic_facts,navistrat_sic_wafer_market_2025} \\

Native GaN &
Bulk-GaN offerings are commonly reported in approximately 2--4-in. formats, with historically limited supplier diversity. &
Supports low-defect vertical-GaN structures and thick high-voltage drift regions, but limited wafer diameter, substrate cost, and supplier concentration constrain near-term volume deployment. &
\cite{osada2017development,universitywafer_bulk_gan,sapphire_substrate_gan_wafer_4inch,yole2017_bulk_gan_market,compoundsemi2012_gan_costs_plummet} \\

Upstream gallium supply &
Recent USGS reporting attributes approximately 99\% of worldwide primary low-purity gallium production to China. &
Geographic concentration introduces procurement, geopolitical, and price-volatility risks. Data-center deployment therefore benefits from multiple qualified suppliers, foundries, and substrate pathways. &
\cite{usgs2025gallium,usgs2026gallium,moon2026china} \\

\midrule
\multicolumn{4}{l}{\textbf{Metallization, packaging, and thermal integration}} \\

Gold compatibility in Si fabs &
Shared-tool contamination-control rules may restrict Au-containing materials and processes. &
Au-free process flows improve compatibility with established Si-fabrication infrastructure and reduce the need for isolated process equipment. &
\cite{cnf_material_compatibility,tanaka_case13,breach2008copper_gold_ball_bonding} \\

Au-to-Cu wire transition &
One industry case study reports material-cost reduction of up to 90\% when replacing Au bonding wire with Cu wire. &
Can reduce interconnect cost and dependence on precious metals, although Cu-wire reliability, bond-pad compatibility, and assembly conditions must be qualified for the selected package. &
\cite{tanaka_case13,strydom2012gallium,reusch2013gan_paralleling} \\

Package inductance &
Representative estimates include less than 0.2~nH for an LGA-class package, approximately 0.5~nH for a QFN Cu-clip package, and approximately 1.5~nH for an SO-8-class package. &
Nanohenry-scale parasitics influence drain overshoot, ringing, switching loss, false turn-on, and EMI under high $\mathrm{d}v/\mathrm{d}t$ and $\mathrm{d}i/\mathrm{d}t$. Package architecture can therefore determine whether the intrinsic switching capability of GaN is usable. &
\cite{reusch2013gan_paralleling,gansystems_gs66508t_datasheet_2020} \\

Package thermal resistance &
Representative datasheets report top-side junction-to-case thermal resistance of approximately 0.28~$^\circ$C/W and a bottom-cooled example near 0.5~$^\circ$C/W. &
The package, die attach, PCB, and cold-plate interface frequently dominate the effective thermal path. Advanced cooling packages improve heat extraction but may increase assembly complexity and yield sensitivity. &
\cite{ti_lmg3522,inf_igt65r055,idtechex_dieattach_ev,trendforceinnoscience} \\

\midrule
\multicolumn{4}{l}{\textbf{Qualification, yield, and lifecycle impact}} \\

HTOL qualification baseline &
A commonly cited qualification reference point is 1000~h of high-temperature operating life at a junction temperature of 125~$^\circ$C or higher, depending on the device and applicable standard. &
Provides a common reliability framework, but lifetime extrapolation is valid only when the acceleration model corresponds to the relevant gate, trapping, interconnect, or package failure mechanism. &
\cite{TI_SNOAA68_2021,Bizo2022SiliconHeatwave,Runyon2023PredictiveAnalyticsDatacenter} \\

Manufacturing yield &
One public report cites a 97\% yield example in a 200-mm GaN manufacturing context. &
Yield affects device cost and the manufacturing burden per functional die. The value should be treated as a process-specific example rather than a universal GaN yield benchmark. &
\cite{TI_SNOAA68_2021,Chen2023LifeTestMILJEDEC,Eikenberg2022AutomotiveReliabilityTesting} \\

Embodied-emissions anchor &
A published wafer-level lifecycle assessment reports direct post-abatement emissions on the order of tens of kilograms of CO$_2$e per 300-mm wafer, with representative cases of approximately 65--88~kg CO$_2$e/wafer.\tnote{a} &
Provides a screening-level basis for examining how die yield, package burden, equipment lifetime, and replacement frequency affect embodied emissions per delivered converter service. &
\cite{de2023search} \\

\bottomrule
\end{tabularx}

\begin{tablenotes}[flushleft]
\footnotesize
\item[a] The wafer-level value is a general semiconductor-manufacturing proxy and should not be interpreted as a GaN-specific lifecycle inventory without process-specific validation.
\end{tablenotes}
\end{threeparttable}
\end{table*}

\subsection{Substrate Platforms and Manufacturing Scalability}

Commercial GaN technologies use three principal substrate pathways: GaN-on-Si, GaN-on-SiC, and native GaN \cite{hsu2021development,kaminski2014sic}. Each pathway produces a different balance among wafer cost, defect density, thermal performance, voltage scalability, process compatibility, and manufacturing volume.

GaN-on-Si currently offers the clearest pathway toward high-volume manufacturing because it can leverage large-diameter silicon wafers and established semiconductor-fabrication infrastructure \cite{dadgar2015sixteen,hsu2021development}. This platform is particularly relevant to lateral devices in the voltage classes used by many data-center PFC, isolated-conversion, and intermediate-bus stages. However, the economic advantage of the Si substrate does not eliminate epitaxial complexity. Lattice and thermal-expansion mismatch require engineered nucleation, transition, and buffer layers to control cracking, residual stress, vertical leakage, and wafer bow \cite{spiridon2021method}. These effects influence process uniformity, usable wafer area, trapping behavior, dynamic $R_{\mathrm{DS(on)}}$, and final device yield.

GaN-on-SiC prioritizes thermal performance and material compatibility relative to GaN-on-Si, supported by the high thermal conductivity of SiC and its more favorable lattice relationship with GaN \cite{jarndal20212,kaminski2014sic}. This can be advantageous in applications where localized heat flux or sustained power density is a primary constraint. However, GaN-on-SiC also inherits exposure to SiC substrate price, wafer-capacity limitations, and competing demand from electric-vehicle, charging, renewable-energy, and grid-conversion markets. It is therefore most appropriate where the additional thermal or electrical performance justifies the higher substrate and supply-chain burden.

Native-GaN substrates provide the low-defect foundation required by many vertical-GaN device concepts, particularly those using thick drift regions and high blocking voltage \cite{paskova2009gan}. Their advantages include reduced lattice mismatch, lower dislocation density, and the ability to form vertical current paths without the highly mismatched buffer structures required on foreign substrates. Nevertheless, bulk-GaN wafer diameter, availability, cost, and supplier diversity remain comparatively limited \cite{huang2024direct}. Native GaN is consequently better positioned as an enabling platform for specialized high-voltage devices than as a near-term replacement for the large-volume lateral GaN-on-Si ecosystem.

Substrate selection therefore cannot be reduced to material performance alone. GaN-on-Si provides manufacturing scale, GaN-on-SiC provides additional thermal headroom, and native GaN supports high-quality vertical structures. The appropriate choice depends on device voltage, current density, thermal boundary, manufacturing volume, qualification requirements, and acceptable cost per converter function.

\subsection{Metallization, Packaging, and Thermal-Electrical Co-Design}

Metallization and packaging determine whether the semiconductor die can be manufactured using high-volume infrastructure and whether its switching capability is preserved in the converter. The transition from Au toward Cu- and Al-based interconnects can reduce material cost and improve compatibility with Si-fabrication environments in which Au contamination is tightly controlled \cite{liu2021evolution,zhong2022review}. Gold-free processing can therefore broaden foundry access and reduce the need for dedicated contamination-controlled modules \cite{liu2021evolution}.

Cu and Al interconnects may also provide favorable electrical and thermal conduction, but their benefits depend on bond-pad metallurgy, intermetallic formation, oxidation control, mechanical stress, and thermal-cycling reliability. Their presence in established recycling streams is a lifecycle advantage, but end-of-life recovery remains dependent on package construction and disassembly economics \cite{lin2020gallium}. Metallization choice should therefore be treated primarily as a manufacturing and reliability decision, with sustainability benefits evaluated as a secondary consequence.

Packaging is particularly consequential for GaN because fast voltage and current transitions make small parasitic elements electrically significant \cite{jorgensen2018fast,wang2022review}. Common-source inductance and power-loop inductance increase gate ringing, drain overshoot, switching energy, and EMI. Parasitic capacitance can increase common-mode displacement current and isolation stress. Low-inductance package formats, Kelvin-source connections, Cu clips, embedded dies, and co-packaged gate drivers can preserve fast-switching performance, but they also increase assembly and qualification complexity.

The thermal path must be co-designed with the electrical package. Advanced substrates, sintered interfaces, top-side cooling, double-sided cooling, and high-conductivity baseplates can reduce junction-to-coolant thermal resistance \cite{wang2022review}. However, increasing package complexity can introduce additional interfaces, coefficient-of-thermal-expansion mismatch, void sensitivity, and assembly-yield loss. A more complex module is advantageous only when its lower loss or longer service life offsets the added material and manufacturing burden. Long-lived, highly integrated modules may therefore provide better lifecycle performance than frequently replaced lower-cost packages, provided that reliability and repairability are considered during design \cite{lumbreras2021effect}.

\subsection{Qualification, Yield, and Lifetime}

Reliability and manufacturing yield link semiconductor technology directly to converter economics and lifecycle impact \cite{kozak2023stability}. A device that offers lower switching loss but exhibits poor yield, unstable dynamic behavior, or shortened package lifetime may increase cost, spare inventory, maintenance requirements, and replacement frequency. These concerns are especially important in data centers, where power systems operate continuously and component failure can compromise service availability.

Conventional qualification tests provide a necessary baseline, but they do not by themselves reproduce the complete converter mission profile. GaN devices may experience repetitive high-voltage off-state stress, rapid hard or soft commutation, reverse conduction, load transients, startup and shutdown events, and elevated local heat flux. Relevant degradation mechanisms include charge trapping, dynamic $R_{\mathrm{DS(on)}}$, gate leakage, threshold-voltage shift, dielectric degradation, metallization fatigue, die-attach degradation, and solder- or interconnect-related thermal cycling \cite{calzolaro2022status,kozak2023stability}.

High-temperature operating life must therefore be supplemented by dynamic and system-representative testing. Qualification should include repetitive switching stress, temperature cycling, power cycling, surge and fault events, current sharing, EMI-related overstress, and operation across the expected partial-load range. Acceleration factors should be linked to a physically relevant failure mechanism rather than applied generically.

Yield improvements reduce the number of wafers, epitaxial runs, packages, and assembly operations required per functional device. Lifetime improvements reduce the number of replacement devices and modules required over the service life of the facility \cite{meneghini2021gan}. Both mechanisms lower cost and embodied burden per unit of delivered power-conversion service. Improvements in epitaxy, passivation, gate-stack design, field management, metallization, and packaging therefore strengthen the economic and lifecycle case simultaneously \cite{meneghini2021gan}.

\subsection{Supply-Chain and Data-Center Deployment Implications}

GaN supply-chain risk extends beyond the availability of elemental gallium. Commercial deployment also depends on epitaxial reactors, qualified wafer suppliers, foundry capacity, package assembly, test infrastructure, and access to second-source devices with sufficiently comparable switching and reliability characteristics \cite{yeboah2025wide,lu2025design}. A nominally equivalent transistor from another supplier may require changes in gate drive, dead time, protection thresholds, layout, thermal interface, or EMI filtering.

For mission-critical data-center infrastructure, second-source qualification must therefore occur at both the component and converter levels. Device procurement strategies should consider package compatibility, dynamic $R_{\mathrm{DS(on)}}$, gate-voltage limits, reverse-conduction behavior, thermal impedance, reliability data, and long-term product availability. Multi-source wafer and foundry strategies can reduce geographic and supplier concentration, but only when the resulting process variants maintain consistent electrical and reliability behavior.

The practical adoption criterion is ultimately total cost of ownership rather than device price alone. A higher-cost GaN device or package may be justified if it reduces converter losses, passive-component volume, cooling demand, rack footprint, or maintenance frequency. Conversely, a device with excellent laboratory performance may provide limited infrastructure value if it requires a single-source substrate, unusually complex assembly, or insufficiently demonstrated lifetime. Manufacturing scale, package performance, qualification maturity, and supply continuity must therefore be evaluated together with efficiency and power density when selecting GaN for data-center power conversion.

\section{Quantitative Energy and Operational Carbon Impact}

While GaN is often justified by device-level efficiency and power-density advantages, a lower-carbon data-center argument requires an explicit link between converter performance, facility electricity consumption, and CO$_2$ emissions \cite{manganelli2021strategies}. This section introduces a compact quantitative framework connecting (i) the multiplicative efficiency of cascaded power-conversion stages, (ii) the incremental cooling demand associated with converter losses, and (iii) embodied impacts associated with manufacturing yield and replacement cycles \cite{yang2022increasing}. The framework provides order-of-magnitude sensitivity estimates for how stage-level efficiency improvements, such as sub-percentage-point gains enabled by GaN in front-end PFC or high-frequency isolated conversion, affect annual facility energy consumption and operational emissions \cite{zhang2022prediction}. The objective is not to predict a universal change in PUE or CO$_2$ emissions, but to establish a transparent, parameterized model that can be populated using site-specific values for load, utilization, cooling performance, grid carbon intensity, and equipment lifetime \cite{mondal2023geeco}.

\begin{figure*}[!t]
\centering
\includegraphics[width=\textwidth]{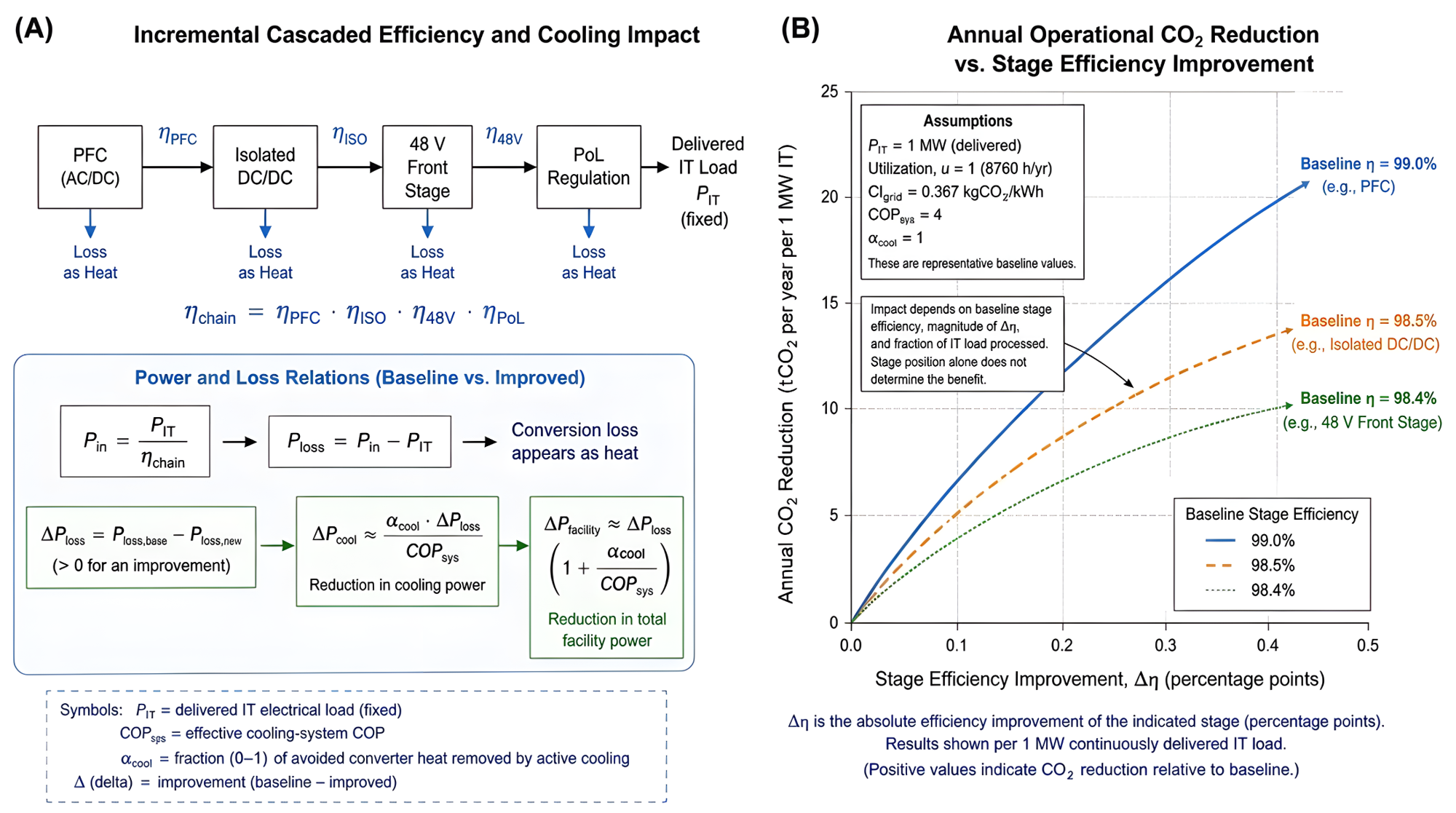}
\caption{(a) Incremental energy-flow model for a representative cascaded data-center power-delivery chain. For a fixed delivered IT load, the stage efficiencies determine the required chain input power and total conversion loss. An efficiency improvement reduces converter loss by $\Delta P_{\mathrm{loss}}$ and produces an associated cooling-power reduction of approximately $\alpha_{\mathrm{cool}}\Delta P_{\mathrm{loss}}/\mathrm{COP}{\mathrm{sys}}$. (b) Illustrative annual operational CO$2$ reduction as a function of absolute stage-efficiency improvement for a 1-MW continuously delivered IT load. Differences among the curves arise from the assumed baseline stage efficiencies and load processed; stage position alone does not determine the system-level benefit. The calculation assumes $u=1$, $\mathrm{COP}{\mathrm{sys}}=4$, $\alpha{\mathrm{cool}}=1$, and $\mathrm{CI}_{\mathrm{grid}}=0.367~\mathrm{kg~CO_2/kWh}$.}
\label{fig:chain_sensitivity}
\end{figure*}

\subsection{System Boundary and Metrics}

The electrical boundary used here separates the power delivered to the IT equipment from the power entering the selected conversion chain. The delivered IT load, $P_{\mathrm{IT}}$, represents the regulated electrical power supplied at the output of the modeled chain. Facility power additionally includes conversion losses, cooling energy, and other overhead loads. Carbon-reduction progress is considered through both (i) use-phase emissions associated with IT and facility electricity consumption and (ii) embodied emissions associated with manufacturing and replacement of power devices and modules \cite{gupta2021chasing}.

Power usage effectiveness is defined as \cite{fieni2025x}

\begin{equation}
\mathrm{PUE}
=
\frac{E_{\mathrm{facility}}}
{E_{\mathrm{IT}}},
\label{eq:pue}
\end{equation}

where $E_{\mathrm{facility}}$ is the total facility energy and $E_{\mathrm{IT}}$ is the energy delivered to IT equipment over the same reporting interval. Carbon usage effectiveness is defined as \cite{liu2025carbon}

\begin{equation}
\mathrm{CUE}
=
\frac{M_{\mathrm{CO_2,facility}}}
{E_{\mathrm{IT}}},
\label{eq:cue}
\end{equation}

where $M_{\mathrm{CO_2,facility}}$ is the operational mass of CO$*2$ associated with facility electricity consumption. For a fully grid-supplied facility using a consistent grid emissions factor, $\mathrm{CI}*{\mathrm{grid}}$, the reporting relationship may be approximated as

\begin{equation}
\mathrm{CUE}
\approx
\mathrm{PUE},
\mathrm{CI}_{\mathrm{grid}}.
\label{eq:cue_pue}
\end{equation}

For avoided-emissions calculations, a marginal grid emissions factor is preferable when available because it represents the generation displaced by the reduction in electricity demand. An average grid factor may instead be used as a transparent screening-level approximation.

Embodied emissions per functional die are represented using the proxy

\begin{equation}
M_{\mathrm{emb,die}}
\approx
\frac{M_{\mathrm{wafer}}}
{N_{\mathrm{die}}Y_{\mathrm{die}}}
+
M_{\mathrm{pkg}}
+
M_{\mathrm{assy}},
\label{eq:embodied_die}
\end{equation}

where $M_{\mathrm{wafer}}$ is the wafer-level embodied greenhouse-gas burden, $N_{\mathrm{die}}$ is the number of gross dies per wafer, $Y_{\mathrm{die}}$ is the functional die yield, and $M_{\mathrm{pkg}}$ and $M_{\mathrm{assy}}$ represent packaging and assembly contributions, respectively. Wafer-level greenhouse-gas data can therefore be propagated through yield, device count, service life, and replacement assumptions \cite{li2023toward}. The numerical example below evaluates operational effects only; embodied impacts should be added separately when sufficiently resolved lifecycle data are available.

\subsection{Cascaded Efficiency and Incremental Cooling Demand}

For a power-conversion chain containing $n$ series-connected stages with efficiencies $\eta_i$, the end-to-end efficiency is multiplicative \cite{monch2023highly}:

\begin{equation}
\eta_{\mathrm{chain}}
=
\prod_{i=1}^{n}\eta_i.
\label{eq:chain_efficiency}
\end{equation}

For a fixed delivered IT load $P_{\mathrm{IT}}$, the electrical power entering the modeled conversion chain is \cite{sesotyo2025evaluating}

\begin{equation}
P_{\mathrm{in}}
=
\frac{P_{\mathrm{IT}}}
{\eta_{\mathrm{chain}}},
\label{eq:input_power}
\end{equation}

and the total power-conversion loss is

\begin{equation}
P_{\mathrm{loss}}
=
P_{\mathrm{in}}
-
P_{\mathrm{IT}}.
\label{eq:conversion_loss}
\end{equation}

For a change in the efficiency of stage $j$, the sensitivity of the chain input power is

\begin{equation}
\frac{\partial P_{\mathrm{in}}}
{\partial \eta_j}
=
-
\frac{P_{\mathrm{IT}}}
{\eta_{\mathrm{chain}}\eta_j}.
\label{eq:stage_sensitivity}
\end{equation}

Equation~\eqref{eq:stage_sensitivity} shows that, for a purely series-connected chain serving the same load, the impact of a stage-efficiency improvement depends on the stage efficiency and the magnitude of its change rather than on its physical position alone. In practical branched power architectures, stages that process a larger fraction of the total IT load provide greater absolute leverage than converters serving only an individual board, processor, or load branch.

Although both the delivered IT power and the conversion losses ultimately appear primarily as heat within the facility thermal boundary \cite{vikhor2024modeling}, the delivered IT load is held constant when comparing the baseline and improved converters. Therefore, the incremental cooling benefit is governed by the reduction in conversion loss rather than by the total IT heat load. Defining the positive loss reduction as

\begin{equation}
\Delta P_{\mathrm{loss}}
=
P_{\mathrm{loss,base}}
-
P_{\mathrm{loss,new}}
=
P_{\mathrm{in,base}}
-
P_{\mathrm{in,new}},
\label{eq:loss_reduction}
\end{equation}

the corresponding reduction in cooling power can be approximated as

\begin{equation}
\Delta P_{\mathrm{cool}}
\approx
\frac{\alpha_{\mathrm{cool}}
\Delta P_{\mathrm{loss}}}
{\mathrm{COP}_{\mathrm{sys}}},
\label{eq:cooling_reduction}
\end{equation}

where $\mathrm{COP}*{\mathrm{sys}}$ is the effective cooling-system coefficient of performance and $0\leq\alpha*{\mathrm{cool}}\leq1$ represents the fraction of avoided converter heat that would otherwise be removed by the active cooling system. For power electronics located entirely within the conditioned thermal boundary, $\alpha_{\mathrm{cool}}\approx1$ provides a useful screening-level case. If a portion of the loss is dissipated outside the actively cooled space, $\alpha_{\mathrm{cool}}<1$ should be used.

The resulting reduction in facility electrical power is

\begin{equation}
\Delta P_{\mathrm{facility}}
\approx
\Delta P_{\mathrm{loss}}
\left(
1+
\frac{\alpha_{\mathrm{cool}}}
{\mathrm{COP}_{\mathrm{sys}}}
\right),
\label{eq:facility_power_reduction}
\end{equation}

assuming that all other facility overheads remain unchanged \cite{le2019techno}. Because Eq.~\eqref{eq:facility_power_reduction} already accounts explicitly for the direct electrical-loss reduction and its cooling consequence, multiplying the result by PUE would double count at least part of the same power and cooling overhead. PUE is therefore retained as a facility-reporting metric and contextual parameter, but it is not used as an additional multiplier in the incremental calculation below.

\subsection{Illustrative Full-Utilization Sensitivity Case}

Representative baseline parameters used in the sensitivity calculation are summarized in Table~\ref{tab:vi_baseline}. These values are illustrative anchors rather than universal data-center operating conditions and should be replaced with site-specific parameters in a detailed assessment.

\begin{table}[!ht]
\caption{Baseline assumptions for the compact sensitivity calculation \cite{ishraq2024design,sun2021mitigation,okilly2022design,lee2016application,chrysostomou2021multicell,reusch2018system,ahmed2020two,pilawa2024extreme}.}
\label{tab:vi_baseline}
\centering
\footnotesize
\setlength{\tabcolsep}{3pt}
\renewcommand{\arraystretch}{1.10}
\begin{tabularx}{\columnwidth}{
@{}
>{\raggedright\arraybackslash}p{0.48\columnwidth}
>{\raggedright\arraybackslash}X
@{}}
\toprule
\textbf{Parameter} & \textbf{Assumed value} \\
\midrule
Delivered IT load, $P_{\mathrm{IT}}$ & 1~MW \\
Annual operating time, $h_{\mathrm{yr}}$ & 8760~h/year \\
Utilization factor, $u$ & 1.0 \\
Grid emissions factor, $\mathrm{CI}_{\mathrm{grid}}$ & 0.81~lb CO$_2$/kWh ($0.367$~kg CO$_2$/kWh) \\
Cooling-system COP, $\mathrm{COP}_{\mathrm{sys}}$ & 4 \\
Actively cooled loss fraction, $\alpha_{\mathrm{cool}}$ & 1.0 \\
PFC efficiency, $\eta_{\mathrm{PFC}}$ & 0.990 baseline; 0.993 improved \\
Isolated-stage efficiency, $\eta_{\mathrm{iso}}$ & 0.985 \\
48-V front-stage efficiency, $\eta_{\mathrm{48V,front}}$ & 0.984 \\
\bottomrule
\end{tabularx}
\end{table}

The calculation assumes a constant 1-MW delivered IT load operating continuously throughout the year. Thus, $u=1$ represents a full-utilization sensitivity case rather than a universal data-center operating profile. The assumed cooling-system coefficient of performance is $\mathrm{COP}{\mathrm{sys}}=4$, and $\alpha{\mathrm{cool}}=1$ assumes that all avoided converter heat would otherwise be removed by the active cooling system. PUE is not applied as an additional multiplier because the incremental cooling contribution is calculated explicitly through $\mathrm{COP}_{\mathrm{sys}}$; applying both factors would risk double counting facility overhead.

Using the three-stage representative chain,
\begin{equation}
\begin{aligned}
\eta_{\mathrm{chain,base}}
&=
0.990
\times
0.985
\times
0.984 \
&=
0.95955.
\end{aligned}
\label{eq:chain_base}
\end{equation}

This multiplicative structure represents a cascaded data-center power-delivery architecture in which the selected conversion stages are connected in series \cite{pilawa2024extreme}. If the PFC efficiency increases from 0.990 to 0.993, corresponding to a 0.3-percentage-point improvement consistent with reported high-performance GaN PFC demonstrations \cite{lee2022design,sun2021mitigation}, the new chain efficiency is
\begin{equation}
\begin{aligned}
\eta_{\mathrm{chain,new}}
&=
0.993
\times
0.985
\times
0.984 \
&=
0.96246.
\end{aligned}
\label{eq:chain_new}
\end{equation}

Although the absolute PFC efficiency improvement is only 0.3 percentage points, the nominal PFC loss fraction decreases from 1.0\% to 0.7\%, corresponding to a 30\% reduction in that stage's fractional loss under the assumed operating condition.

For $P_{\mathrm{IT}}=1~\mathrm{MW}$, the reduction in input power is

\begin{equation}
\begin{aligned}
\Delta P_{\mathrm{in}}
&=
P_{\mathrm{IT}}
\left(
\frac{1}{\eta_{\mathrm{chain,base}}}
-
\frac{1}{\eta_{\mathrm{chain,new}}}
\right) \\
&=
1~\mathrm{MW}
\left(
\frac{1}{0.95955}
-
\frac{1}{0.96246}
\right) \\
&\approx
3.15~\mathrm{kW}.
\end{aligned}
\label{eq:input_reduction_example}
\end{equation}

Because the delivered IT load is unchanged,

\begin{equation}
\Delta P_{\mathrm{loss}}
=
\Delta P_{\mathrm{in}}
\approx
3.15~\mathrm{kW}.
\label{eq:loss_reduction_example}
\end{equation}

For $\mathrm{COP}*{\mathrm{sys}}=4$ and $\alpha*{\mathrm{cool}}=1$, the facility-level power reduction becomes

\begin{equation}
\begin{aligned}
\Delta P_{\mathrm{facility}}
&=
3.15
\left(
1+
\frac{1}{4}
\right) \
&\approx
3.94~\mathrm{kW}.
\end{aligned}
\label{eq:facility_reduction_example}
\end{equation}

This result includes both the direct reduction in converter electricity consumption and the incremental reduction in cooling power associated with the avoided heat load \cite{huang2023cooling,chang2024optimization}.

For an annual utilization factor $u$, the corresponding energy reduction is

\begin{equation}
\Delta E_{\mathrm{annual}}
=
\Delta P_{\mathrm{facility}},
u,
h_{\mathrm{yr}}.
\label{eq:annual_energy_reduction}
\end{equation}

Using $u=1$ and $h_{\mathrm{yr}}=8760~\mathrm{h/year}$,

\begin{equation}
\begin{aligned}
\Delta E_{\mathrm{annual}}
&=
3.94~\mathrm{kW}
\times
8760~\mathrm{h/year} \
&\approx
34.5~\mathrm{MWh/year}.
\end{aligned}
\label{eq:annual_energy_example}
\end{equation}

The associated operational emissions reduction is

\begin{equation}
\Delta M_{\mathrm{CO_2}}
=
\Delta E_{\mathrm{annual}},
\mathrm{CI}_{\mathrm{grid}}.
\label{eq:carbon_reduction}
\end{equation}

Using $\mathrm{CI}_{\mathrm{grid}}=0.367~\mathrm{kg~CO_2/kWh}$ gives

\begin{equation}
\Delta M_{\mathrm{CO_2}}
\approx
12.7~\mathrm{tCO_2/year}
\label{eq:carbon_example}
\end{equation}

per 1~MW of continuously delivered IT load under the stated assumptions.

\subsection{Interpretation and Model Limitations}

The result in Eq.~\eqref{eq:carbon_example} is an illustrative full-utilization sensitivity case rather than a universal annual savings estimate. Actual facility performance depends on the converter efficiency curves over the operating range, server utilization, redundancy configuration, cooling architecture, climate, airflow management, location of the power electronics relative to the conditioned thermal boundary, and temporal variation in grid carbon intensity.

In particular, peak converter efficiency should not be assumed to represent annual average efficiency. AI training, inference, networking, and storage workloads can produce time-varying electrical demand, and redundant power systems may operate individual modules below their rated load. A more detailed model should therefore use time-indexed load and efficiency data,

\begin{equation}
\Delta E_{\mathrm{annual}}
=
\sum_{k}
\Delta P_{\mathrm{facility},k}
\Delta t_k,
\label{eq:time_resolved_energy}
\end{equation}

or an equivalent integral over the annual load profile.

Similarly, the stage providing the largest benefit is not necessarily the earliest stage in the chain. In a series-connected model, all efficiencies multiply, and the sensitivity depends on the baseline efficiency and magnitude of the improvement. In an actual data center, absolute leverage also depends on the fraction of facility power processed by that converter. A front-end PFC stage may therefore have high aggregate leverage because it serves an entire power shelf or rack, whereas a point-of-load converter serves only a particular load branch, not simply because the PFC stage appears earlier in the conversion chain \cite{pilawa2024extreme}.

Finally, the operational result does not by itself establish carbon neutrality. A complete lifecycle assessment must also account for embodied emissions, manufacturing yield, equipment lifetime, replacement frequency, refrigerants, onsite generation, storage, electricity procurement, and end-of-life treatment. The framework instead provides a traceable method for translating a measured stage-efficiency improvement into an incremental facility-energy and operational-emissions consequence without conflating PUE, cooling COP, and converter loss.

\section{Deployment Pathways and Design Guidelines}
\label{sec:deployment_guidelines}

GaN adoption in data-center power systems should be treated as a staged system-integration problem rather than as a direct transistor substitution \cite{ma2019review}. The attainable benefit depends on placing GaN in conversion stages where switching and commutation losses represent a meaningful fraction of total loss and where increased switching frequency can reduce magnetic and capacitive volume without creating excessive electromagnetic-interference (EMI), thermal, or reliability penalties \cite{wang20243,li2023survey}. Device selection must therefore be coordinated with converter topology, package, gate drive, layout, protection, magnetics, and cooling.

\begin{figure*}[!t]
\centering
\includegraphics[width=\linewidth]{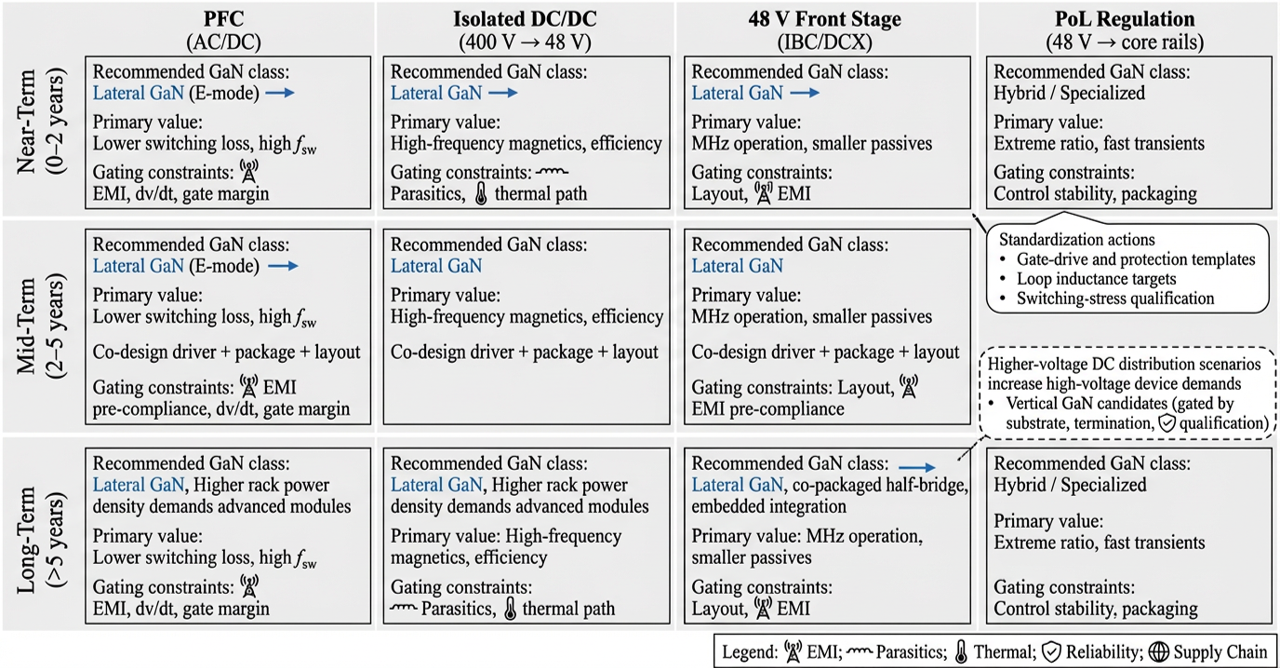}
\caption{Staged deployment roadmap for GaN devices across representative data-center PFC, isolated DC/DC, 48-V front-stage, and PoL conversion functions. Near-term opportunities emphasize commercially mature lateral GaN and demonstrated converter topologies; mid-term opportunities require tighter device--package--driver--thermal co-design; and long-term opportunities include emerging high-voltage, bidirectional, and highly integrated architectures. The roadmap indicates expected deployment readiness rather than a fixed implementation schedule.}
\label{fig:gan_deployment_roadmap}
\end{figure*}

This section organizes the deployment pathway by technology maturity and power-chain function. Fig.~\ref{fig:gan_deployment_roadmap} identifies representative near-, mid-, and long-term opportunities across power-factor-correction (PFC), isolated DC/DC, 48-V intermediate conversion, and point-of-load (PoL) regulation. Fig.~\ref{fig:gan_design_flow} complements this roadmap by illustrating the coupled design variables that determine whether GaN's intrinsic device capability produces a repeatable converter-level benefit.

\subsection{Near-Term Deployment in Mature Converter Stages}

Near-term adoption should prioritize converter stages for which lateral GaN devices, suitable topologies, and supporting package technologies are already commercially available or extensively demonstrated. These stages include front-end PFC, resonant isolated conversion from a high-voltage DC link to a 48-V-class bus, and selected fixed-ratio intermediate-bus converters \cite{lee2022design}. These applications process a large fraction of rack or power-shelf throughput and operate under switching conditions in which commutation behavior, output capacitance, reverse conduction, and magnetic-component size strongly affect converter performance.

Open Compute Project Open Rack V3 specifies a nominal 48-V distribution architecture with an IT-equipment input range of 46--52~V. Centralized rack- or shelf-level conversion followed by 48-V distribution reduces conductor current and corresponding $I^{2}R$ loss relative to lower-voltage rack buses. It also moves the final high-ratio conversion closer to the processor or accelerator load. In the front-end and isolated stages, lateral GaN can reduce switching and commutation losses while supporting higher operating frequency, smaller magnetics, and increased volumetric power density \cite{lee2016application}.

Near-term deployment should follow three priorities. First, enhancement-mode lateral GaN should be applied in totem-pole PFC and resonant isolated DC/DC topologies where its low switching charge and absence of conventional body-diode reverse recovery provide a direct system advantage \cite{lee2022design}. The switching frequency should be selected from a total-loss optimum rather than from the maximum capability of the device, because increased frequency also raises magnetic, gate-drive, capacitive, and EMI-related losses.

Second, package and PCB layout should be treated as first-order electrical design variables. Commutation-loop and gate-loop inductances should be estimated before hardware fabrication and verified through impedance extraction, switching-waveform measurements, or both. Device replacement without corresponding loop and driver redesign is unlikely to realize the expected GaN performance.

Third, qualification should include application-relevant repetitive switching and transient conditions in addition to conventional static-bias tests. The test envelope should reproduce representative $\mathrm{d}v/\mathrm{d}t$, $\mathrm{d}i/\mathrm{d}t$, reverse-conduction intervals, line and load transients, startup, shutdown, and thermal cycling. These operating conditions can govern field reliability even when the nominal voltage, current, and junction-temperature ratings are not exceeded.

\subsection{Mid-Term Device--Package--Driver Co-Design}

Mid-term deployment is characterized by increasing rack power, higher converter density, and closer integration of power conversion with server boards and accelerator packages. Under these conditions, the limiting factors progressively shift from the intrinsic transistor die toward package parasitics, driver placement, current sharing, thermal interfaces, integrated magnetics, and EMI compliance \cite{chrzan2025gan}. As indicated in Fig.~\ref{fig:gan_deployment_roadmap}, this phase requires coordinated design of the semiconductor, package, driver, PCB, magnetic components, protection system, and cooling interface.

Gate-drive practices should be standardized around the selected GaN architecture rather than transferred directly from silicon MOSFET platforms. Important design variables include turn-on and turn-off resistance, gate-voltage margin, source inductance, local driver decoupling, negative turn-off bias where permitted, and active control of abnormal switching events. Miller-clamp or equivalent false-turn-on mitigation, fast overcurrent detection, and coordinated soft shutdown may be required depending on the device and topology \cite{bau2020cmos}. Protection response must be sufficiently fast for the limited fault-energy tolerance of high-power-density GaN devices, while avoiding nuisance triggering during normal high-$\mathrm{d}i/\mathrm{d}t$ operation.

EMI becomes a primary deployment constraint as edge rates and switching frequencies increase into the hundreds-of-kilohertz and megahertz ranges \cite{jia2024mitigating}. An explicit $\mathrm{d}v/\mathrm{d}t$ and $\mathrm{d}i/\mathrm{d}t$ strategy is therefore required to balance switching loss, overshoot, common-mode current, and conducted and radiated emissions. Relevant measures include minimizing commutation-loop area, separating power and gate return paths, using Kelvin-source connections where available, controlling switch-node copper area, optimizing common-mode capacitance, and selecting packages with low parasitic inductance.

\begin{figure*}[!t]
\centering
\includegraphics[width=\linewidth]{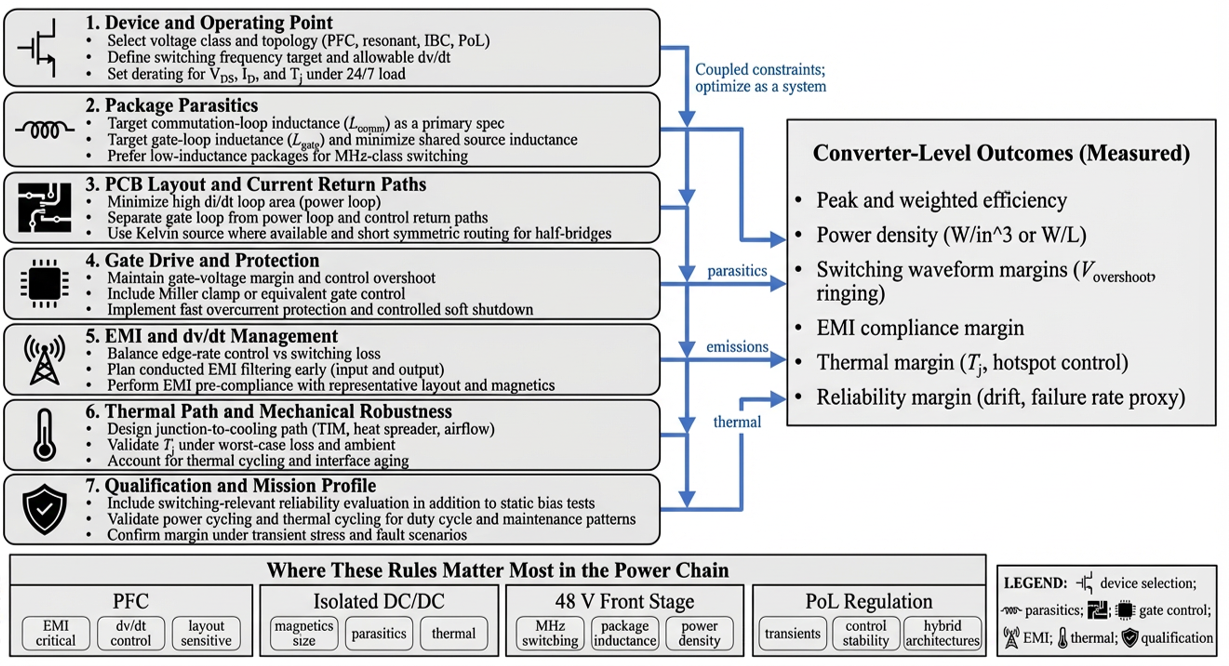}
\caption{System-level GaN converter design workflow showing the coupling among device selection, package and interconnect parasitics, PCB layout, gate drive and protection, EMI management, magnetic design, and thermal pathways. Converter efficiency, power density, waveform quality, compliance, and reliability emerge from the combined design rather than from the semiconductor device alone.}
\label{fig:gan_design_flow}
\end{figure*}

EMI pre-compliance measurements should begin with early converter prototypes rather than after efficiency and density targets have been finalized. Otherwise, additional filters, shielding, slower gate drive, or layout changes may eliminate the expected power-density advantage. The coupling among parasitics, switching waveforms, emissions, and converter performance is summarized in Fig.~\ref{fig:gan_design_flow}.

Thermal and mechanical reliability also become more restrictive as the converter footprint decreases. Lower total loss does not necessarily imply a lower junction temperature when the same loss is concentrated into a smaller package area. The thermal path from the active region through the package, PCB or substrate, interface material, heat spreader, and cooling system must therefore be evaluated using realistic spatial heat flux and boundary conditions. Qualification should include power cycling and thermal cycling representative of data-center load changes, redundancy operation, maintenance events, and environmental excursions \cite{kozak2023stability}.

\subsection{Long-Term High-Voltage and Integrated Architectures}

Long-term deployment is associated with more substantial changes in data-center power architecture, including higher-voltage DC distribution, greater converter modularity, bidirectional energy interfaces, solid-state protection, and increased integration of power stages with rack- and board-level loads \cite{rothmund201899}. These architectures may alter both the voltage class and functional requirements imposed on GaN devices.

Several-hundred-volt DC buses, including $\pm 400$-V and 800-V-class concepts, are being considered as possible methods for reducing distribution current, conductor mass, and busbar loss as rack power increases. Such architectures also create additional requirements for insulation coordination, DC fault interruption, protection selectivity, galvanic isolation, and local high-ratio conversion. Their deployment should therefore be treated as prospective rather than as an established replacement for 48-V rack distribution.

Higher distribution voltages may increase the relevance of vertical GaN devices, advanced edge termination, monolithic bidirectional switches, and series-connected modular conversion. However, blocking voltage alone is not sufficient to justify adoption. Native-substrate quality, defect density, edge termination, package isolation, surge capability, short-circuit behavior, and high-voltage qualification must reach infrastructure-grade maturity \cite{meneghini2021gan}. Until then, lateral GaN may remain applicable through modular converter cells, while SiC continues to provide an established alternative in high-voltage and high-power stages.

Highly integrated long-term architectures may combine GaN switches, drivers, protection, sensing, and control within a common package or power integrated circuit. Integration can reduce gate-loop inductance and improve switching reproducibility, but it can also increase local heat flux, common-mode coupling, repair difficulty, and dependence on a single package platform. The resulting value must therefore be assessed at the converter and rack levels rather than from component count alone.

Long-term adoption also requires manufacturing and qualification milestones. Multi-source substrate, foundry, and package strategies are needed to limit supply-chain concentration. Reliability reporting should include switching-relevant stress and package wear-out under realistic temperature and power cycling. More transparent cradle-to-gate lifecycle data are also required to determine whether increased package complexity and manufacturing burden are offset by longer service life, reduced operational loss, and lower replacement frequency \cite{manganelli2021strategies}.

\subsection{Stage-Specific Design Guidelines}

Independent of deployment horizon, the following guidelines increase the probability that GaN integration produces a repeatable system-level advantage. They reflect the coupled dependencies in Fig.~\ref{fig:gan_design_flow}, where the device, package, gate drive, layout, magnetics, EMI control, protection, and thermal path jointly determine converter performance \cite{chrzan2025gan,prajapati2023leveraging}.

\textbf{1) Select stages according to loss mechanism and processed power.}
GaN should be prioritized where switching, output-capacitance, reverse-conduction, or commutation losses constitute a meaningful part of total converter loss. The absolute infrastructure benefit also depends on the fraction of rack or facility load served by the stage. A front-end stage may provide high aggregate leverage because it processes an entire power shelf or rack, not simply because it appears earlier in the conversion chain.

\textbf{2) Optimize switching frequency at the converter level.}
The switching frequency should minimize total system loss and volume rather than semiconductor switching loss alone. Device loss, magnetic core and winding loss, capacitor loss, gate-drive power, EMI-filter volume, and cooling requirements should be evaluated together.

\textbf{3) Specify package and layout limits quantitatively.}
Targets should be established for commutation-loop inductance, gate-loop inductance, common-source inductance, switch-node capacitance, and thermal impedance. These quantities should be verified using extraction, impedance measurement, double-pulse testing, or converter-level waveform analysis \cite{sun2019research}. The package must be treated as part of the electrical and thermal circuit rather than as a mechanical enclosure.

\textbf{4) Maintain gate-drive and protection margins.}
Gate voltage, drain-voltage overshoot, reverse-conduction interval, and peak current should be verified over device tolerance, temperature, load, and parasitic variation. Protection thresholds and response times should be selected for the fault withstand capability of the specific GaN architecture \cite{bau2020cmos}.

\textbf{5) Co-design EMI and efficiency.}
Switching-edge control should balance transition loss against common-mode current, ringing, overshoot, and filter requirements. EMI testing should use representative magnetics, cables, grounding, enclosure geometry, and cooling hardware because laboratory open-bench waveforms may not reproduce the final emissions environment \cite{derkacz2022emi}.

\textbf{6) Evaluate partial-load and transient performance.}
Peak efficiency alone is insufficient for continuously operated data-center infrastructure. Converter efficiency, temperature, regulation, and current sharing should be characterized across the expected load distribution, redundancy state, and accelerator transient profile.

\textbf{7) Qualify the complete mission profile.}
Static-bias life testing should be supplemented by repetitive switching, reverse conduction, thermal cycling, power cycling, surge, startup, shutdown, overload, and fault-recovery tests. Switching-reliability guidance for GaN power devices should be applied together with converter-specific stress analysis \cite{derkacz2022emi}.

\subsection{Deployment Decision Framework}

A practical deployment decision should compare the measurable system benefit against the implementation penalties introduced by the new technology. GaN adoption is justified when reductions in semiconductor loss, passive-component volume, distribution loss, and cooling burden outweigh the additional requirements for packaging, gate drive, EMI control, protection, qualification, and supply-chain management.

This comparison can be expressed conceptually as

\begin{equation}
\begin{split}
\Delta B_{\mathrm{system}}
=\; &\Delta B_{\mathrm{efficiency}}
+
\Delta B_{\mathrm{density}}
+
\Delta B_{\mathrm{thermal}} \\
&-\,\Delta C_{\mathrm{integration}}
-
\Delta C_{\mathrm{qualification}},
\end{split}
\label{eq:deployment_balance}
\end{equation}

where the benefit terms represent improvements in conversion efficiency, power density, and thermal burden, while the cost terms represent additional integration and qualification requirements. The quantities need not share a common physical unit at the initial screening stage, but a final deployment decision should translate them into total cost of ownership, energy consumption, rack capacity, failure risk, or another system-level objective.

Accordingly, the recommended pathway is heterogeneous rather than universal: mature lateral GaN in near-term high-frequency stages, tighter device--package--driver integration in mid-term high-density conversion, and vertical, bidirectional, or highly integrated GaN only where their manufacturing and qualification maturity supports the required infrastructure reliability. These guidelines translate the material and device advantages developed in Sections~\ref{sec:materials} and \ref{sec:device_architecture_landscape} into a practical deployment framework for next-generation AI data centers.

\section{Open Challenges and Research Gaps}
\label{sec:open_challenges}

Despite substantial progress in GaN materials, device architectures, packaging, and converter demonstrations, several coupled barriers continue to limit fleet-scale deployment in mission-critical data centers \cite{islam2022reliability,lavrivc2025challenges}. These barriers extend beyond intrinsic device physics and arise from interactions among switching stress, package parasitics, electrothermal behavior, manufacturing variability, and the reliability requirements of continuous operation \cite{qin2023thermal,wang2023review}. Resolving them requires measurement and qualification methods that reproduce converter mission profiles rather than relying only on static device ratings or best-case laboratory operating points.

\begin{figure*}[!t]
\centering
\includegraphics[width=\linewidth]{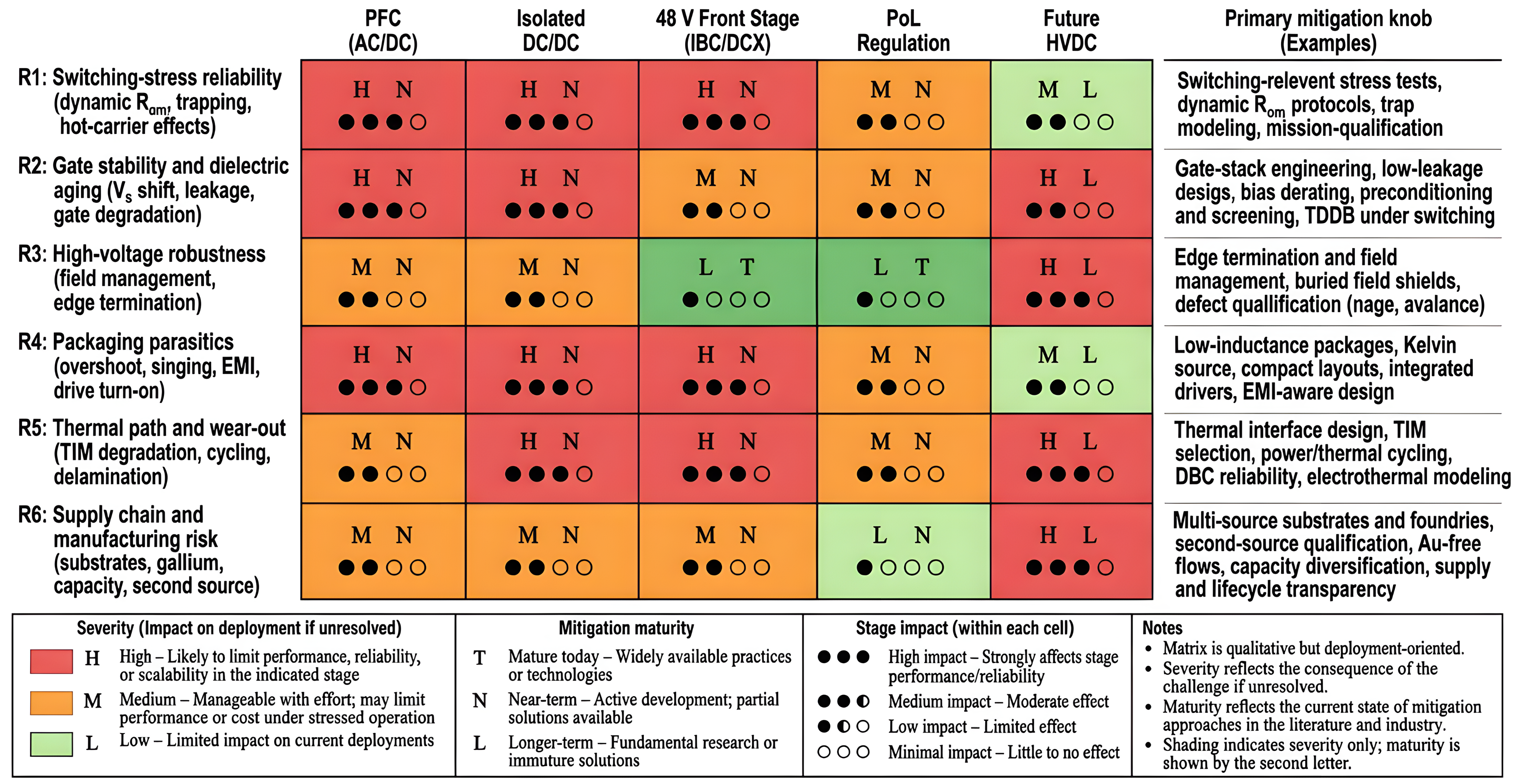}
\caption{Qualitative matrix of deployment-limiting challenges for GaN data-center power conversion. Rows identify device, packaging, qualification, benchmarking, and supply-chain constraints, while columns represent representative stages of the data-center power chain. The first letter in each cell denotes estimated severity (H: high, M: medium, L: low), and the second denotes mitigation maturity (T: available today, N: requiring near-term development, L: requiring longer-term research). Cell shading represents severity only. The classifications indicate relative research priority and should not be interpreted as quantitative failure probabilities.}
\label{fig:gan_challenges_matrix}
\end{figure*}

Fig.~\ref{fig:gan_challenges_matrix} qualitatively summarizes the principal deployment constraints across representative stages of the data-center power chain. The matrix distinguishes the estimated severity of each barrier from the maturity of available mitigation approaches. It should be interpreted as a research-priority map rather than as a quantitative failure-risk assessment.

\subsection{Switching-Stress Reliability: Dynamic On-Resistance, Trapping, and Gate Stability}

Charge trapping and field-driven degradation remain central reliability concerns for lateral GaN power HEMTs operated under realistic switching conditions \cite{kozak2023stability,meneghini2021gan}. Trapped charge in surface, barrier, buffer, or interface states can increase dynamic on-resistance, produce current collapse, and alter threshold voltage after high-voltage off-state stress. The resulting loss increase may not be apparent from static datasheet values, yet it can raise junction temperature and change converter efficiency during repetitive operation \cite{kozak2023stability}.

A central interpretation challenge is that reversible trapping and permanent degradation can occur simultaneously. Dynamic $R_{\mathrm{DS(on)}}$ may partially recover after the stress is removed, while defect generation, gate degradation, or package aging may produce cumulative changes. Measurements taken after different recovery delays can therefore yield substantially different results even for the same device and stress condition. Comparisons among published studies are further complicated by differences in drain-voltage stress, switching waveform, temperature, duty cycle, current level, measurement aperture, and preconditioning.

Normally-off devices introduce additional stability concerns. In p-GaN gate HEMTs, carrier injection and defect dynamics in the gate region can modify gate current and threshold voltage during positive gate-bias stress \cite{islam2022reliability}. Extracted $V_{\mathrm{th}}$ values can also depend on the measurement method, sweep direction, delay time, and prior bias history. A reported threshold shift must therefore be distinguished from a measurement-induced transient or reversible charge redistribution.

Three research priorities follow from these limitations. First, dynamic $R_{\mathrm{DS(on)}}$ protocols should specify the off-state stress voltage, switching waveform, junction temperature, stress duration, recovery delay, and measurement aperture. Second, threshold-voltage and gate-current measurements should use controlled preconditioning and separate reversible instability from permanent aging. Third, device-level stress results should be connected to converter-level consequences, including additional conduction loss, temperature rise, current-sharing error, and control-margin reduction.

Structure-level approaches that reduce gate leakage, stabilize the threshold voltage, and suppress trapping remain important research directions \cite{meneghini2021gan}. Switching-oriented reliability guidance, including JEDEC JEP180.01, also reinforces the need to supplement conventional qualification with application-relevant dynamic testing \cite{JEDEC2021JEP180,Kozak2023GaNReview,Tayyab2022Dynamic}. The unresolved deployment question is not simply whether a device passes a static lifetime test, but whether its switching parameters remain sufficiently stable throughout the expected converter mission profile \cite{Kozak2023GaNReview,Rauf2025Investigation,Tayyab2022Dynamic,Fan2024Dynamic}.

\subsection{High-Voltage and Vertical GaN: Field Management, Ruggedness, and Edge Termination}

Vertical GaN offers a potential pathway toward higher blocking voltage and improved voltage-area scaling, but its deployment introduces additional high-field and process-integration challenges \cite{langpoklakpam2023vertical,mukherjee2021challenges}. In trench-gate devices, electric-field crowding near dielectric corners can accelerate dielectric degradation or cause premature breakdown during off-state operation \cite{mukherjee2021challenges}. Field plates, buried field shields, optimized trench geometry, and dielectric-stack engineering can reduce peak field, but their effectiveness depends strongly on dimensional tolerances, interface quality, charge distribution, and process reproducibility.

Edge termination is similarly critical in high-voltage vertical diodes and transistors \cite{langpoklakpam2023vertical}. The active region may support a high theoretical breakdown field, while the practical device fails at a substantially lower voltage because of field crowding at the die perimeter, implantation damage, surface charge, or crystallographic defects. Termination effectiveness must therefore be evaluated together with leakage stability, temperature dependence, process variability, and long-duration high-field stress \cite{mukherjee2021challenges}.

Further research is required in three areas. First, high-voltage structures need termination and dielectric-protection designs that remain effective across realistic fabrication tolerances rather than only at an optimized nominal geometry. Second, reported breakdown voltage should be accompanied by leakage-current behavior, temperature dependence, active area, termination dimensions, and statistical device-to-device variation. Third, ruggedness metrics—including surge response, repetitive overvoltage, fault behavior, and short-circuit tolerance—must be characterized under conditions relevant to high-voltage DC distribution and infrastructure converters.

The principal deployment gap is therefore the transition from isolated high-voltage demonstrations to manufacturable, yield-robust devices with package insulation and qualification evidence appropriate for mission-critical systems. Until those requirements are met, vertical GaN should be treated as a high-potential but emerging platform rather than as a mature replacement for established high-voltage technologies.

\subsection{Packaging-Limited Scaling: Parasitics, Thermal Interfaces, and Cycling Wear-Out}

Packaging often determines whether the intrinsic switching capability of GaN produces lower converter loss or instead produces excessive overshoot, ringing, false turn-on, and EMI \cite{wang2024advanced,wohrle2023power}. Common-source inductance, gate-loop inductance, power-loop inductance, and package capacitance influence switching energy and voltage stress. The resulting behavior depends on the complete commutation path, including the transistor package, driver placement, PCB layout, decoupling network, and magnetic-component connections \cite{wang2024advanced}.

Thermal behavior presents a parallel limitation. High-frequency GaN converters may reduce total loss while concentrating the remaining dissipation into a smaller die and package area. Junction temperature is therefore governed by local heat flux and the stability of the full thermal path rather than by efficiency alone. Die attach, substrate metallization, direct-bonded-copper structures, interface materials, PCB vias, heat spreaders, and cold plates can each become lifetime-limiting interfaces.

Thermal-cycling studies of low-inductance GaN modules have shown that package degradation can appear as an increase in thermal resistance, including delamination and interface deterioration that may not be detected through switching-waveform monitoring alone \cite{sun2024thermal}. This creates a need for coupled electrical, thermal, and mechanical diagnostics. Useful indicators include thermal-impedance evolution, structure-function analysis, transient thermal response, switching-energy drift, and post-stress imaging or failure analysis.

Future package research should therefore target simultaneous reductions in electrical parasitics and thermal impedance while preserving manufacturability and cycling lifetime. Qualification should reproduce both rapid electrical transients and slower thermal-mechanical cycling. Package comparisons should also report the cooling boundary, interface material, mounting pressure, substrate configuration, and measured parasitics so that device-level advantages can be separated from package-level effects.

A further challenge is balancing integration with lifecycle performance. Advanced modules may reduce operating loss and cooling demand, but excessive material complexity, low assembly yield, difficult repair, or package-driven wear-out can weaken the overall sustainability benefit \cite{yadlapalli2021advancements}. Package optimization must therefore include lifetime, repairability, and manufacturing yield in addition to switching performance.

\subsection{Manufacturing Variability and Supply-Chain Resilience}

Large-scale GaN deployment depends on stable access to substrate materials, epitaxial capacity, foundry processes, package assembly, and upstream critical materials. Public supply assessments indicate substantial geographic concentration in parts of the gallium and compound-semiconductor value chain \cite{wesselkaemper2025enhancing}. This concentration can expose converter programs to price volatility, export controls, long qualification cycles, and limited second-source options.

For mission-critical data centers, second sourcing cannot be based only on nominal voltage and current ratings. Devices from different suppliers may differ in threshold voltage, gate-voltage margin, output capacitance, reverse-conduction behavior, dynamic $R_{\mathrm{DS(on)}}$, package inductance, and thermal impedance. Substituting a nominally equivalent device may therefore require changes in gate drive, dead time, overcurrent protection, EMI filtering, and cooling.

Research and manufacturing priorities include qualification of multiple substrate and wafer sources, process flows compatible with high-volume Si manufacturing, and device designs tolerant of substrate and epitaxial variability \cite{zhong2022review}. Statistical reporting of wafer-level uniformity, dynamic behavior, yield, and reliability distribution would also improve the ability to assess supply-chain substitutability.

Lifecycle claims require similar transparency. Process-specific cradle-to-gate inventories, package-material data, equipment lifetime, and end-of-life recovery assumptions remain limited. Better data are required to determine whether operational energy savings offset the embodied burden of advanced substrates and packages \cite{yeboah2025wide}. The resulting research need is a combined technical and supply-chain qualification framework that evaluates electrical equivalence, reliability equivalence, manufacturing continuity, and lifecycle burden.

\subsection{Benchmarking and Reporting Gaps}

Published GaN results are difficult to compare when device and converter metrics are reported under inconsistent operating conditions. Device studies may emphasize minimum static $R_{\mathrm{DS(on)}}$, low switching energy, or maximum breakdown voltage, while infrastructure deployment depends on efficiency and stability across load, temperature, switching frequency, cooling, EMI, and transient conditions. Review literature continues to identify dynamic $R_{\mathrm{DS(on)}}$, gate degradation, and electrothermal accumulation as central barriers requiring improved defect-level and system-level understanding \cite{kozak2023stability}.

At the device level, useful reporting should include voltage and current waveforms, junction or case temperature, gate resistance, loop inductance, switching speed, reverse-conduction interval, dead time, and the timing used to extract dynamic parameters. At the converter level, reports should specify topology, device part number and voltage class, switching mode, switching frequency, magnetic design, package and layout, cooling boundary, peak and full-load efficiency, partial-load efficiency, power density, EMI conditions, and protection strategy.

Reliability studies should report the complete stress waveform and distinguish among reversible trapping, parametric drift, catastrophic failure, and package degradation. Where accelerated lifetime models are used, the assumed failure mechanism and acceleration law should be stated explicitly. Converter studies should also identify whether demonstrated efficiency was measured at nominal, peak, or thermally stabilized operation and whether auxiliary, gate-drive, fan, and control power were included.

A standardized benchmark for data-center GaN converters should therefore include:

\begin{enumerate}
\item the fraction of rack or facility power processed by the stage;
\item complete efficiency curves over the expected load and temperature range;
\item semiconductor, magnetic, interconnect, and auxiliary loss breakdowns;
\item extracted or measured package and commutation-loop parasitics;
\item thermal boundary conditions and stabilized device temperatures;
\item conducted and radiated EMI results or defined pre-compliance conditions; and
\item mission-profile reliability testing under representative switching and cycling stress.
\end{enumerate}

Such reporting would allow device, package, and topology improvements to be compared on a common converter-level basis rather than through isolated best-case metrics.

\subsection{Priority Research Directions}

The most important research gaps cluster around five coupled objectives: switching-stress stability, high-field ruggedness of vertical devices, package electrothermal lifetime, manufacturing and supply-chain reproducibility, and standardized converter benchmarking. These categories and their stage-specific implications are summarized in Fig.~\ref{fig:gan_challenges_matrix}.

Near-term research should prioritize standardized dynamic-parameter measurement, package-parasitic characterization, mission-profile qualification, and comparable converter reporting. Mid-term research should target integrated electrothermal package design, stable normally-off gate structures, multi-source process qualification, and partial-load optimization. Longer-term research must establish manufacturable vertical-GaN structures, high-voltage package insulation, bidirectional and protection functionality, and process-specific lifecycle inventories.

Addressing these gaps is necessary to move GaN from successful component and converter demonstrations to repeatable deployment across large data-center fleets. The decisive criterion is not whether GaN can achieve a record efficiency at one operating point, but whether the complete device--package--converter system can sustain its electrical, thermal, and reliability advantages throughout manufacturing variation and the intended service life.

\section{Conclusion}

Gallium nitride provides a broad device and converter design space for addressing the increasing efficiency, power-density, and thermal demands of AI data-center power delivery. Its value, however, is stage dependent rather than universal. Commercially mature lateral GaN HEMTs are particularly well suited to high-frequency PFC, resonant isolated DC/DC, and intermediate-bus conversion, where switching and commutation losses strongly influence converter performance. Specialized and hybrid architectures extend this capability to normally-off control, bidirectional power flow, extreme conversion ratios, and functional integration. Vertical GaN offers a potential pathway toward higher-voltage and higher-power conversion, but its deployment remains dependent on continued progress in substrate availability, edge termination, dielectric protection, packaging, and infrastructure-grade qualification.

At the system level, GaN performance cannot be inferred from material properties or device figures of merit alone. Converter benefit emerges from coordinated selection of the device architecture, topology, switching frequency, package, gate drive, magnetics, layout, protection, EMI strategy, and thermal boundary. When these elements are co-optimized, GaN can reduce conversion loss, increase volumetric power density, and lower the heat generated within the power-delivery chain. The associated facility-energy and operational-carbon benefit depends on the magnitude of the efficiency improvement, the fraction of rack power processed by the affected stage, converter utilization, cooling-system performance, and grid carbon intensity. GaN should therefore be regarded as an enabling technology for lower-energy and lower-carbon operation rather than as a device substitution that independently guarantees a fixed reduction in PUE or emissions.

The principal requirement for broader deployment is no longer the demonstration of isolated peak-efficiency records, but the achievement of repeatable device--package--converter performance under realistic mission profiles, manufacturing variation, and long service life. Progress is particularly needed in switching-stress stability, dynamic $R_{\mathrm{DS(on)}}$ characterization, gate reliability, high-field ruggedness of vertical devices, package electrothermal lifetime, multi-source manufacturing qualification, and standardized converter benchmarking. Addressing these challenges through coordinated device, packaging, converter, and infrastructure development will determine whether GaN becomes a foundational technology for efficient, high-density, and reliable AI data-center power systems.




%

\bibliographystyle{IEEEtran}
\bibliography{References/Bibliography}

@article{matsunami2020fundamental,
  title={Fundamental research on semiconductor SiC and its applications to power electronics},
  author={Matsunami, Hiroyuki},
  journal={Proceedings of the Japan Academy, Series B},
  volume={96},
  number={7},
  pages={235--254},
  year={2020},
  publisher={The Japan Academy}
}

@phdthesis{ayalew2004sic,
  title={SiC semiconductor devices technology, modeling and simulation},
  author={Ayalew, Tesfaye},
  year={2004},
  school={Technische Universit{\"a}t Wien}
}

@article{matocha2008challenges,
  title={Challenges in SiC power MOSFET design},
  author={Matocha, Kevin},
  journal={Solid-State Electronics},
  volume={52},
  number={10},
  pages={1631--1635},
  year={2008},
  publisher={Elsevier}
}

@article{zhang2025applicability,
  title={Applicability Analysis of High-Voltage Transmission and Substation Equipment Based on Silicon Carbide Devices},
  author={Zhang, Huiyuan and Nie, Ming and Dong, Qinxiao and Liu, He and Jia, Pengfei and Li, Zhiyuan and Fang, Yonghao},
  journal={Micromachines},
  volume={16},
  number={11},
  pages={1192},
  year={2025},
  publisher={MDPI}
}

@article{dai2026study,
  title={Study on the grinding mechanism of single-crystal GaN based on atomic scale and stress field modeling},
  author={Dai, Houfu and Song, Tiantian and Du, Hao},
  journal={Materials Science in Semiconductor Processing},
  volume={204},
  pages={110313},
  year={2026},
  publisher={Elsevier}
}

@techreport{OpenComputeORv3,
  title        = {Open Rack V3 IT Gear 48V Input Connector, Rev. 1.6},
  author       = {{Open Compute Project}},
  institution  = {Open Compute Project},
  type         = {Technical Specification},
  number       = {Rev. 1.6},
  year         = {2022},
  month        = sep,
  url          = {https://bit.ly/40N0xKk},
  note         = {Published 29~September~2022}
}

@techreport{TEORv3Power,
  title        = {OCP ORv3 Power Solutions Guide},
  author       = {{TE Connectivity}},
  institution  = {TE Connectivity},
  type         = {Technical Guide},
  year         = {2023},
  month        = may,
  url          = {https://bit.ly/3MWUakE},
  note         = {Guide for Open Compute Project Open Rack V3 power solutions (10 pages)} 
}

@inproceedings{yu2024novel,
  title={A Novel Digital Control Strategy for GaN-based Interleaving CrM Totem-pole PFC},
  author={Yu, Wenhao and Fan, Xuefeng and Wei, Tao and Xu, Yingchun},
  booktitle={2024 IEEE Energy Conversion Congress and Exposition (ECCE)},
  pages={2831--2836},
  year={2024},
  organization={IEEE}
}

@techreport{InfineonGSEVB,
  title        = {GS-EVB-BTP-3KW-GS Evaluation Board Technical Documentation},
  author       = {{Infineon Technologies AG}},
  institution  = {Infineon Technologies AG},
  type         = {Evaluation Board / Technical Manual},
  year         = {2024},
  url          = {https://www.infineon.com/evaluation-board/GS-EVB-BTP-3KW-GS},
  note         = {3 kW High-Efficiency Bridgeless Totem Pole PFC evaluation board (CoolGaN family) and supporting technical manual},
}

@techreport{EPCbp092025,
  title        = {Wide Bandgap Power Conversion – Part 3: ISOP LLC Converter},
  author       = {Alejandro Pozo and Michael de Rooij and Marco Palma},
  institution  = {Efficient Power Conversion (EPC)},
  type         = {Technical Article},
  year         = {2025},
  month        = sep,
  url          = {https://epc-co.com/epc/portals/0/epc/documents/articles/bp_092025.pdf},
  note         = {Published 28 September 2025 as part of Bodo’s Power Systems series},
}

@inproceedings{ahmed2019gan,
  title={GaN based high-density unregulated 48 V to x V LLC converters with??? 98\% efficiency for future data centers},
  author={Ahmed, Mohamed H and Lee, Fred C and Li, Qiang and de Rooij, Michael and Reusch, David},
  booktitle={PCIM Europe 2019; International Exhibition and Conference for Power Electronics, Intelligent Motion, Renewable Energy and Energy Management},
  pages={1--8},
  year={2019},
  organization={VDE}
}

@inproceedings{baek2020lego,
  title={LEGO-PoL: A 48V-1.5 V 300A merged-two-stage hybrid converter for ultra-high-current microprocessors},
  author={Baek, Jaeil and Wang, Ping and Elasser, Youssef and Chen, Yenan and Jiang, Shuai and Chen, Minjie},
  booktitle={2020 IEEE Applied Power Electronics Conference and Exposition (APEC)},
  pages={490--497},
  year={2020},
  organization={IEEE}
}

@article{shehabi20242024,
  title={2024 united states data center energy usage report},
  author={Shehabi, Arman and Newkirk, Alex and Smith, Sarah J and Hubbard, Alex and Lei, Nuoa and Siddik, Md Abu Bakar and Holecek, Billie and Koomey, Jonathan and Masanet, Eric and Sartor, Dale},
  year={2024}
}

@techreport{UptimeInstitute2024,
  title        = {Uptime Institute Global Data Center Survey 2024},
  author       = {{Uptime Institute}},
  institution  = {Uptime Institute},
  type         = {Technical Report},
  number       = {UII Keynote Report 146M},
  year         = {2024},
  month        = jul,
  url          = {https://datacenter.uptimeinstitute.com/rs/711-RIA-145/images/2024.GlobalDataCenterSurvey.Report.pdf},
  note         = {Findings highlight resiliency, sustainability, efficiency, staffing, cloud, and AI trends among data center owners and operators}
}

@techreport{ASHRAETC92016,
  title        = {{Data Center Power Equipment Thermal Guidelines and Best Practices}},
  author       = {{ASHRAE Technical Committee 9.9}},
  institution  = {ASHRAE},
  type         = {White Paper},
  year         = {2016},
  month        = jun,
  url          = {https://www.ashrae.org/file%20library/technical%20resources/bookstore/ashrae_tc0909_power_white_paper_22_june_2016_revised.pdf},
  note         = {Revised 22 June 2016; thermal guidelines for data center power equipment}
}

@online{universitywafer_silicon_wafers,
  author       = {{UniversityWafer, Inc.}},
  title        = {Silicon Wafers},
  year         = {n.d.},
  url          = {https://www.universitywafer.com/silicon_wafers.html},
  note         = {Accessed: 2026-03-03}
}

@online{universitywafer_gan_on_si_epitaxy,
  author       = {{UniversityWafer, Inc.}},
  title        = {Gallium Nitride on Silicon Epitaxial Wafers},
  year         = {n.d.},
  url          = {https://www.universitywafer.com/gallium-nitride-on-silicon-epitaxy-wafer.html},
  note         = {Accessed: 2026-03-03}
}

@online{okmetic_rf_gan_substrate_wafers,
  author       = {{Okmetic}},
  title        = {RF GaN Substrate Wafers for GaN-on-Si Applications},
  year         = {n.d.},
  url          = {https://www.okmetic.com/silicon-wafers/rfsi-wafers-high-resistivity-line-for-rf/rf-gan-substrate-wafers/},
  note         = {Accessed: 2026-03-03}
}

@online{navitas_200mm_gan_psmc_2025,
  author       = {{Navitas Semiconductor}},
  title        = {Navitas Announces Plans for 200mm GaN Production with PSMC},
  year         = {2025},
  month        = jul,
  day          = {1},
  url          = {https://bit.ly/4bmGOHy},
  note         = {Accessed: 2026-03-03}
}

@online{xfab_gan_on_si_foundry_2025,
  author       = {{X-FAB Silicon Foundries SE}},
  title        = {X-FAB Now Offers GaN-on-Si Foundry Services},
  year         = {2025},
  month        = sep,
  day          = {2},
  url          = {https://www.xfab.com/news/details/article/x-fab-now-offers-gan-on-si-foundry-services},
  note         = {Accessed: 2026-03-03}
}

@online{infineon_gan_technology,
  author       = {{Infineon Technologies AG}},
  title        = {Gallium Nitride (GaN) Technology},
  year         = {n.d.},
  url          = {https://www.infineon.com/technology/gallium-nitride-gan},
  note         = {Accessed: 2026-03-03}
}

@online{infineon_300mm_gan_press_2024,
  author       = {{Infineon Technologies AG}},
  title        = {Infineon Pioneers World’s First 300 mm Power Gallium Nitride (GaN) Technology},
  year         = {2024},
  month        = sep,
  day          = {11},
  url          = {https://www.infineon.com/press-release/2024/infxx202409-142},
  note         = {Accessed: 2026-03-03}
}

@misc{gansystems_gs66508t_datasheet_2020,
  author       = {{GaN Systems Inc.}},
  title        = {GS66508T: 650 V Enhancement Mode GaN Transistor Datasheet, Rev. 200402},
  year         = {2020},
  url          = {https://www.mouser.com/datasheet/3/70/1/GS66508T-DS-Rev-200402.pdf},
  note         = {Accessed: 2026-03-03}
}

@techreport{coherent_sic_materials_datasheet,
  author       = {{Coherent Corp.}},
  title        = {Silicon Carbide (SiC) Materials},
  institution  = {Coherent Corp.},
  year         = {n.d.},
  type         = {Datasheet},
  url          = {https://www.coherent.com/resources/datasheet/materials/sic-materials-ds.pdf},
  note         = {Accessed: 2026-03-03}
}

@online{trendforce_sic_price_war_2025,
  author       = {{TrendForce}},
  title        = {SiC Raw Materials See a Price Increase While 6-Inch Substrate Kicks off a Price War},
  year         = {2025},
  date         = {2025-11-26},
  url          = {https://bit.ly/4aNZCzj},
  organization = {TrendForce},
  note         = {Accessed: 2026-03-03}
}

@online{navitas_sic_facts,
  author       = {{Navitas Semiconductor}},
  title        = {Silicon Carbide: The Facts},
  year         = {n.d.},
  organization = {Navitas Semiconductor},
  url          = {https://navitassemi.com/silicon-carbide-the-facts/},
  note         = {Accessed: 2026-03-03}
}

@online{navistrat_sic_wafer_market_2025,
  author       = {{Navistrat Analytics}},
  title        = {Silicon Carbide (SiC) Wafer Market Size, Share and Trend Analysis, Forecast 2025--2032},
  year         = {2025},
  organization = {Navistrat Analytics},
  url          = {https://navistratanalytics.com/report_store/silicon-carbide-sic-wafer-market/},
  note         = {Report code: NA\_01409. Accessed: 2026-03-03}
}

@inproceedings{osada2017development,
  title={Development of Non-Core 4-inch GaN Substrate},
  author={Osada, H and Yoshizumi, Y and Uematsu, K and Minobe, S and Sato, F and Nakanisihi, F and Yamamoto, Y and Hagi, Y and Yabuhara, Y},
  booktitle={Proc. Int. Conf. Compound Semiconductor Manufacturing Technology},
  pages={16--4},
  year={2017}
}

@misc{universitywafer_bulk_gan,
  author       = {{UniversityWafer, Inc.}},
  title        = {Bulk Gallium Nitride (GaN) Wafers for Research and Production},
  year         = {n.d.},
  url          = {https://www.universitywafer.com/bulk-gan.html},
  note         = {Accessed: 2026-03-03}
}

@misc{sapphire_substrate_gan_wafer_4inch,
  author       = {{Sapphire-Substrate.com}},
  title        = {4 Inch Research Grade 0.4 mm Free Standing GaN Wafer for Semiconductors},
  year         = {n.d.},
  url          = {https://bit.ly/4aQ1MOU},
  note         = {Accessed: 2026-03-03}
}

@article{yole2017_bulk_gan_market,
  title   = {Bulk GaN substrate market growing at 10\% CAGR to \$100m in 2022, from 60,000 wafers in 2016},
  journal = {semiconductorTODAY Compounds \& Advanced Silicon},
  year    = {2017},
  volume  = {12},
  number  = {2},
  month   = {March/April},
  url     = {https://www.semiconductor-today.com/features/PDF/semiconductor-today-mar-apr-2017-bulk-gan.pdf},
  note    = {Market focus: GaN materials}
}

@article{compoundsemi2012_gan_costs_plummet,
  title   = {Gallium Nitride Substrate Costs to Plummet by 60 Percent},
  journal = {Compound Semiconductor},
  year    = {2012},
  month   = nov,
  day     = {6},
  url     = {https://compoundsemiconductor.net/article/90117/Gallium_nitride_substrate_costs_to_plummet_by_60_percent},
  note    = {Accessed: 2026-03-03}
}

@techreport{usgs2025gallium,
  author       = {{U.S. Geological Survey}},
  title        = {Gallium},
  year         = {2025},
  institution  = {U.S. Geological Survey},
  series       = {Mineral Commodity Summaries},
  month        = jan,
  url          = {https://pubs.usgs.gov/periodicals/mcs2025/mcs2025-gallium.pdf},
  note         = {In: Mineral Commodity Summaries 2025}
}

@techreport{usgs2026gallium,
  author       = {{U.S. Geological Survey}},
  title        = {Gallium},
  year         = {2026},
  institution  = {U.S. Geological Survey},
  series       = {Mineral Commodity Summaries},
  number       = {2026},
  address      = {Reston, VA},
  url          = {https://pubs.usgs.gov/periodicals/mcs2026/mcs2026-gallium.pdf},
  note         = {In: Mineral Commodity Summaries 2026}
}

@techreport{moon2026china,
  author       = {Moon, Ji Won},
  title        = {The Mineral Industry of China},
  institution  = {U.S. Geological Survey},
  series       = {Minerals Yearbook},
  volume       = {III},
  number       = {Area Reports---International},
  year         = {2026},
  note         = {2023 Minerals Yearbook, advance release},
  address      = {Reston, VA},
  url          = {https://pubs.usgs.gov/myb/vol3/2023/myb3-2023-china.pdf}
}

@misc{cnf_material_compatibility,
  author       = {{Cornell NanoScale Science \& Technology Facility}},
  title        = {Material Compatibility},
  year         = {n.d.},
  howpublished = {\url{https://www.cnf.cornell.edu/equipment/compatibility}},
  note         = {Accessed: 2026-03-03},
  institution  = {Cornell University}
}

@misc{tanaka_case13,
  author       = {{TANAKA Precious Metals}},
  title        = {High Precision, High Durability, Low Cost Bead Dishes},
  year         = {n.d.},
  howpublished = {\url{https://tanaka-preciousmetals.com/en/solution/case/case13/}},
  note         = {Accessed: 2026-03-03},
  institution  = {TANAKA Precious Metal Group}
}

@article{breach2008copper_gold_ball_bonding,
  author  = {Christopher Breach},
  title   = {The Great Debate: Copper vs. Gold Ball Bonding},
  journal = {Semiconductor Digest},
  year    = {2008},
  month   = {October},
  url     = {https://sst.semiconductor-digest.com/2008/10/the-great-debate-copper-vs-gold-ball-bonding/},
  note    = {Accessed: 2026-03-03}
}

@article{strydom2012gallium,
  title={Gallium nitride transistor packaging advances and thermal modeling},
  author={Strydom, Johan and de Rooij, Michael and Lidow, Alex},
  journal={EDN China},
  pages={1--13},
  year={2012}
}

@techreport{reusch2013gan_paralleling,
  author       = {David Reusch and Johan Strydom},
  title        = {Effectively Paralleling Enhancement Mode Gallium Nitride Transistors for High Current and High Frequency Applications},
  institution  = {Efficient Power Conversion Corporation (EPC)},
  type         = {Application Note},
  number       = {AN020},
  year         = {2013},
  month        = {May},
  url          = {https://epc-co.com/epc/portals/0/epc/documents/application-notes/AN020%20Effectively%20Paralleling%20Enhancement%20Mode%20Gallium%20Nitride%20Transistors.pdf},
  note         = {Accessed: 2026-03-03}
}

@misc{ti_lmg3522,
  title        = {LMG3522R030-Q1 650-V 30-m$\Omega$ GaN FET With Integrated Driver, Protection, and Temperature Reporting},
  author       = {{Texas Instruments}},
  year         = {2024},
  note         = {Rev. D datasheet},
  url          = {https://www.ti.com/lit/ds/symlink/lmg3522r030-q1.pdf}
}

@misc{inf_igt65r055,
  title        = {IGT65R055D2 CoolGaN™ G5 650 V Enhancement-Mode Power Transistor},
  author       = {{Infineon Technologies AG}},
  year         = {2024},
  note         = {Datasheet, Revision 0.1},
  url          = {https://www.infineon.com/assets/row/public/documents/24/49/infineon-igt65r055d2-datasheet-en.pdf}
}

@misc{idtechex_dieattach_ev,
  title        = {Die Attach Materials for Power Electronics in Electric Vehicles 2020--2030},
  author       = {{IDTechEx}},
  year         = {2020},
  url          = {https://bit.ly/4b4E3JC},
  note         = {Market research report}
}

@online{trendforceinnoscience,
  title        = {Innoscience GaN Products Break Into Google's Supply Chain},
  author       = {{TrendForce}},
  year         = {2026},
  date         = {2026-02-09},
  url          = {https://bit.ly/3OEa5F0},
  note         = {News article}
}

@techreport{TI_SNOAA68_2021,
  title        = {Achieving GaN Products With Lifetime Reliability},
  author       = {{Texas Instruments}},
  institution  = {Texas Instruments},
  year         = {2021},
  month        = jun,
  number       = {SNOAA68},
  type         = {White Paper},
  url          = {https://www.ti.com/lit/wp/snoaa68/snoaa68.pdf},
  note         = {Accessed: 2026-03-03}
}

@techreport{Bizo2022SiliconHeatwave,
  author       = {Bizo, Daniel},
  title        = {Silicon Heatwave: The Looming Change in Data Center Climates},
  institution  = {Uptime Institute Intelligence},
  type         = {UI Intelligence Report},
  number       = {74},
  year         = {2022},
  month        = aug,
  url          = {https://bit.ly/46YmzxB},
  note         = {Accessed: 2026-03-03}
}

@online{Runyon2023PredictiveAnalyticsDatacenter,
  author       = {Runyon, Wendi},
  title        = {An Industry’s Journey to System-Level Predictive Analytics for the Data Center Starts Now},
  year         = {2023},
  month        = sep,
  day          = {28},
  organization = {Schneider Electric},
  url          = {https://bit.ly/4cmeWEw},
  note         = {Schneider Electric Blog. Accessed: 2026-03-03}
}

@misc{Chen2023LifeTestMILJEDEC,
  author       = {Chen, Yuan},
  title        = {Statistical Interpretation of Life Test: Comparison Between MIL and JEDEC Requirements},
  year         = {2023},
  month        = jun,
  howpublished = {Presentation, NASA Electronic Parts and Packaging (NEPP) Electronics Technology Workshop (ETW)},
  institution  = {NASA Langley Research Center},
  url          = {https://nepp.nasa.gov/docs/etw/2023/15-JUN-THU/1000_Chen_20230009005.pdf},
  note         = {NASA Technical Reports Server Document ID: 20230009005. Accessed: 2026-03-03}
}

@techreport{Eikenberg2022AutomotiveReliabilityTesting,
  author       = {Eikenberg, Thomas},
  title        = {Automotive Electronics Reliability Testing Starts and Ends with the Mission Profile},
  institution  = {Monolithic Power Systems},
  year         = {2022},
  month        = may,
  number       = {Article \#0061, Rev. 1.0},
  url          = {https://media.monolithicpower.com/mps_cms_document/2/0/2020-aip-automotive-electronics-reliability-testing_r1.0.pdf},
  note         = {Accessed: 2026-03-03}
}

@article{de2023search,
  title={Search for new particles at the ILC},
  author={de Vera, Mar{\'\i}a Teresa N{\'u}{\~n}ez Pardo},
  journal={arXiv preprint arXiv:2311.00525},
  year={2023}
}

@article{ji2021ridge,
  title={Ridge-channel AlGaN/GaN normally-off high-electron mobility transistor based on epitaxial lateral overgrowth},
  author={Ji, Xiaoli and Fariza, Aqdas and Zhao, Jie and Wang, Maojun and Wang, Junxi and Yang, Fuhua and Li, Jinmin and Wei, Tongbo},
  journal={Semiconductor Science and Technology},
  volume={36},
  number={7},
  pages={075003},
  year={2021},
  publisher={IOP Publishing}
}

@article{meneghini2017technology,
  title={Technology and reliability of normally-off GaN HEMTs with p-type gate},
  author={Meneghini, Matteo and Hilt, Oliver and Wuerfl, Joachim and Meneghesso, Gaudenzio},
  journal={Energies},
  volume={10},
  number={2},
  pages={153},
  year={2017},
  publisher={MDPI}
}

@inproceedings{kinzer2022advancing,
  title={Advancing GaN Power ICs: Efficiency, Reliability \& Autonomy},
  author={Kinzer, Dan},
  booktitle={2022 24th European Conference on Power Electronics and Applications (EPE'22 ECCE Europe)},
  pages={1--2},
  year={2022},
  organization={IEEE}
}

@article{udabe2023gallium,
  title={Gallium nitride power devices: a state of the art review},
  author={Udabe, Ander and Baraia-Etxaburu, Igor and Diez, David Garrido},
  journal={IEEE Access},
  volume={11},
  pages={48628--48650},
  year={2023},
  publisher={IEEE}
}

@article{rafin2023power,
  title={Power electronics revolutionized: A comprehensive analysis of emerging wide and ultrawide bandgap devices},
  author={Rafin, SM Sajjad Hossain and Ahmed, Roni and Haque, Md Asadul and Hossain, Md Kamal and Haque, Md Asikul and Mohammed, Osama A},
  journal={Micromachines},
  volume={14},
  number={11},
  pages={2045},
  year={2023},
  publisher={MDPI}
}

@article{boschee2024comments,
  title={Comments: Grabbing the Brass Ring To Power the Demand for Data Centers and Generative AI},
  author={Boschee, Pam},
  journal={Journal of Petroleum Technology},
  volume={76},
  number={05},
  pages={8--9},
  year={2024},
  publisher={OnePetro}
}

@article{shankar2024enhancing,
  title={Enhancing Energy Efficiency in AI-Powered Data Centers: Challenges and Solutions},
  author={Shankar, Sahana},
  journal={Marine Biology Research at Bahamas},
  pages={93},
  year={2024}
}

@article{langpoklakpam2023vertical,
  title={Vertical GaN MOSFET power devices},
  author={Langpoklakpam, Catherine and Liu, An-Chen and Hsiao, Yi-Kai and Lin, Chun-Hsiung and Kuo, Hao-Chung},
  journal={Micromachines},
  volume={14},
  number={10},
  pages={1937},
  year={2023},
  publisher={MDPI}
}

@article{he20221,
  title={1.3 kV vertical GaN-based trench MOSFETs on 4-inch free standing GaN wafer},
  author={He, Wei and Li, Jian and Liao, Zeliang and Lin, Feng and Wu, Junye and Wang, Bing and Wang, Maojun and Liu, Nan and Chiu, Hsien-Chin and Kuo, Hao-Chung and others},
  journal={Nanoscale Research Letters},
  volume={17},
  number={1},
  pages={14},
  year={2022},
  publisher={Springer}
}

@article{duan20231,
  title={1.7-kV vertical GaN pn diode with triple-zone graded junction termination extension formed by ion-implantation},
  author={Duan, Yu and Wang, Jingshan and Xie, Andy and Zhu, Zhongtao and Fay, Patrick},
  journal={e-Prime-Advances in Electrical Engineering, Electronics and Energy},
  volume={6},
  pages={100330},
  year={2023},
  publisher={Elsevier}
}

@article{kaminski2024vertical,
  title={Vertical GaN Trench-MOSFETs Fabricated on Ammonothermally Grown Bulk GaN Substrates},
  author={Kami{\'n}ski, Maciej and Taube, Andrzej and Tarenko, Jaroslaw and Sadowski, Oskar and Brzozowski, Ernest and Wierzbicka, Justyna and Zadura, Magdalena and Ekielski, Marek and Kosiel, Kamil and Jankowska-{\'S}liwi{\'n}ska, Joanna and others},
  journal={physica status solidi (a)},
  volume={221},
  number={21},
  pages={2400077},
  year={2024},
  publisher={Wiley Online Library}
}

@article{ma2025low,
  title={Low leakage fully-vertical GaN-on-Si power MOSFETs},
  author={Ma, Yuchuan and Chen, Hang and Zhang, Shuhui and Duan, Huantao and Hu, Bin and Ma, Huimei and Rao, Jin and Liu, Chao},
  journal={Applied Physics Letters},
  volume={127},
  number={5},
  year={2025},
  publisher={AIP Publishing}
}

@article{islam2022reliability,
  title={Reliability, applications and challenges of GaN HEMT technology for modern power devices: A review},
  author={Islam, Naeemul and Mohamed, Mohamed Fauzi Packeer and Khan, Muhammad Firdaus Akbar Jalaludin and Falina, Shaili and Kawarada, Hiroshi and Syamsul, Mohd},
  journal={Crystals},
  volume={12},
  number={11},
  pages={1581},
  year={2022},
  publisher={MDPI}
}

@article{wang2023review,
  title={Review on main gate characteristics of p-type GaN gate high-electron-mobility transistors},
  author={Wang, Zhongxu and Nan, Jiao and Tian, Zhiwen and Liu, Pei and Wu, Yinhe and Zhang, Jincheng},
  journal={Micromachines},
  volume={15},
  number={1},
  pages={80},
  year={2023},
  publisher={MDPI}
}

@article{alharbi2021experimental,
  title={Experimental evaluation of medium-voltage cascode gallium nitride (GaN) devices for bidirectional DC-DC converters},
  author={Alharbi, Salah S and Matin, Mohammad},
  journal={CES Transactions on Electrical Machines and Systems},
  volume={5},
  number={3},
  pages={232--248},
  year={2021},
  publisher={CES}
}

@article{kozak2023stability,
  title={Stability, reliability, and robustness of GaN power devices: A review},
  author={Kozak, Joseph Peter and Zhang, Ruizhe and Porter, Matthew and Song, Qihao and Liu, Jingcun and Wang, Bixuan and Wang, Rudy and Saito, Wataru and Zhang, Yuhao},
  journal={IEEE Transactions on Power Electronics},
  volume={38},
  number={7},
  pages={8442--8471},
  year={2023},
  publisher={IEEE}
}

@article{cittanti2022new,
  title={New FOM-based performance evaluation of 600/650 V SiC and GaN semiconductors for next-generation EV drives},
  author={Cittanti, Davide and Vico, Enrico and Bojoi, Iustin Radu},
  journal={IEEE Access},
  volume={10},
  pages={51693--51707},
  year={2022},
  publisher={IEEE}
}

@article{musumeci2023gallium,
  title={Gallium nitride power devices in power electronics applications: State of art and perspectives},
  author={Musumeci, Salvatore and Barba, Vincenzo},
  journal={Energies},
  volume={16},
  number={9},
  pages={3894},
  year={2023},
  publisher={MDPI}
}

@article{nguyen2021piezotronic,
  title={Piezotronic effect in a normally off p-GaN/AlGaN/GaN HEMT toward highly sensitive pressure sensor},
  author={Nguyen, Hong-Quan and Nguyen, Thanh and Tanner, Philip and Nguyen, Tuan-Khoa and Foisal, Abu Riduan Md and Fastier-Wooller, Jarred and Nguyen, Tuan-Hung and Phan, Hoang-Phuong and Nguyen, Nam-Trung and Dao, Dzung Viet},
  journal={Applied Physics Letters},
  volume={118},
  number={24},
  year={2021},
  publisher={AIP Publishing}
}

@article{chen2024research,
  title={Research on a High-Threshold-Voltage AlGaN/GaN HEMT with P-GaN Cap and Recessed Gate in Combination with Graded AlGaN Barrier Layer: Z. Chen et al.},
  author={Chen, Zhichao and Cai, Lie and Niu, Kai and Xu, Chaozhi and Lin, Haoxiang and Ren, Pengpeng and Sun, Dong and Lin, Haifeng},
  journal={Journal of Electronic Materials},
  volume={53},
  number={5},
  pages={2533--2543},
  year={2024},
  publisher={Springer}
}

@article{kang2020charging,
  title={Charging effect by fluorine-treatment and recess gate for enhancement-mode on algan/gan high electron mobility transistors},
  author={Kang, Soo Cheol and Jung, Hyun-Wook and Chang, Sung-Jae and Kim, Seung Mo and Lee, Sang Kyung and Lee, Byoung Hun and Kim, Haecheon and Noh, Youn-Sub and Lee, Sang-Heung and Kim, Seong-Il and others},
  journal={Nanomaterials},
  volume={10},
  number={11},
  pages={2116},
  year={2020},
  publisher={MDPI}
}

@article{sunkara2025power,
  title={Power consumption and heat dissipation in AI data centers: A comparative analysis},
  author={Sunkara, Krishnaiah Narukulla Krishna Chaitanya},
  year={2025}
}

@article{li2025ai,
  title={AI Load Dynamics--A Power Electronics Perspective},
  author={Li, Yuzhuo and Li, Yunwei},
  journal={arXiv preprint arXiv:2502.01647},
  year={2025}
}

@article{chaudhary2023technology,
  title={Technology and applications of wide bandgap semiconductor materials: current state and future trends},
  author={Chaudhary, Omar Sarwar and Dena{\"\i}, Mouloud and Refaat, Shady S and Pissanidis, Georgios},
  journal={Energies},
  volume={16},
  number={18},
  pages={6689},
  year={2023},
  publisher={MDPI}
}

@article{meneghini2021gan,
  title={GaN-based power devices: Physics, reliability, and perspectives},
  author={Meneghini, Matteo and De Santi, Carlo and Abid, Idriss and Buffolo, Matteo and Cioni, Marcello and Khadar, Riyaz Abdul and Nela, Luca and Zagni, Nicol{\`o} and Chini, Alessandro and Medjdoub, Farid and others},
  journal={Journal of Applied Physics},
  volume={130},
  number={18},
  year={2021},
  publisher={AIP Publishing}
}

@article{yadlapalli2021advancements,
  title={Advancements in energy efficient GaN power devices and power modules for electric vehicle applications: A review},
  author={Yadlapalli, Ravindranath Tagore and Kotapati, Anuradha and Kandipati, Rajani and Balusu, Srinivasa Rao and Koritala, Chandra Sekhar},
  journal={International Journal of Energy Research},
  volume={45},
  number={9},
  pages={12638--12664},
  year={2021},
  publisher={Wiley Online Library}
}

@article{gupta2022vertical,
  title={Vertical GaN and vertical Ga2O3 power transistors: status and challenges},
  author={Gupta, Chirag and Pasayat, Shubhra S},
  journal={physica status solidi (a)},
  volume={219},
  number={7},
  pages={2100659},
  year={2022},
  publisher={Wiley Online Library}
}

@article{jaiswal2023optimized,
  title={An optimized vertical GaN parallel split gate trench MOSFET device structure for improved switching performance},
  author={Jaiswal, Nilesh Kumar and Ramakrishnan, VN},
  journal={IEEE access},
  volume={11},
  pages={46998--47006},
  year={2023},
  publisher={IEEE}
}

@article{chrzan2025gan,
  title={GaN Power Transistors in Converter Design Techniques},
  author={Chrzan, Piotr J and Derkacz, Pawel B},
  journal={Energies},
  volume={18},
  number={11},
  pages={2890},
  year={2025},
  publisher={MDPI}
}

@article{zhang2023status,
  title={Status and prospects of wide bandgap semiconductor devices},
  author={Zhang, Meihe and Zhang, Yunsong},
  journal={Applied and Computational Engineering},
  volume={23},
  pages={252--262},
  year={2023}
}

@article{shi2023deep,
  title={A deep dive into SiC and GaN power devices: Advances and prospects},
  author={Shi, J},
  journal={Appl. Comput. Eng},
  volume={23},
  number={1},
  pages={230--237},
  year={2023}
}

@article{ma2019review,
  title={Review of GaN HEMT applications in power converters over 500 W},
  author={Ma, Chao-Tsung and Gu, Zhen-Huang},
  journal={Electronics},
  volume={8},
  number={12},
  pages={1401},
  year={2019},
  publisher={MDPI}
}

@inproceedings{rahman2023review,
  title={Review of isolated dc-dc converters for applications in data center power delivery},
  author={Rahman, Syed and Shehada, Halah and Khan, Irfan Ahmad},
  booktitle={2023 IEEE Texas Power and Energy Conference (TPEC)},
  pages={1--6},
  year={2023},
  organization={IEEE}
}

@article{keshmiri2020current,
  title={Current status and future trends of GaN HEMTs in electrified transportation},
  author={Keshmiri, Niloufar and Wang, Deqiang and Agrawal, Bharat and Hou, Ruoyu and Emadi, Ali},
  journal={IEEE access},
  volume={8},
  pages={70553--70571},
  year={2020},
  publisher={IEEE}
}

@article{frivaldsky2020evaluation,
  title={Evaluation of GaN power transistor switching performance on characteristics of bidirectional DC-DC converter},
  author={Frivaldsky, Michal and Morgos, Jan and Zelnik, Richard},
  journal={Elektronika ir elektrotechnika},
  volume={26},
  number={4},
  pages={18--24},
  year={2020}
}

@article{faizan2023design,
  title={Design and comparative analysis of an ultra-highly efficient, compact half-bridge LLC resonant GaN converter for low-power applications},
  author={Faizan, Muhammad and Wang, Xiaolei and Yousaf, Muhammad Zain},
  journal={Electronics},
  volume={12},
  number={13},
  pages={2850},
  year={2023},
  publisher={MDPI}
}

@article{buffolo2024review,
  title={Review and outlook on GaN and SiC power devices: Industrial state-of-the-art, applications, and perspectives},
  author={Buffolo, M and Favero, D and Marcuzzi, A and De Santi, C and Meneghesso, G and Zanoni, E and Meneghini, M},
  journal={IEEE Transactions on Electron Devices},
  volume={71},
  number={3},
  pages={1344--1355},
  year={2024},
  publisher={IEEE}
}

@article{yeboah2025wide,
  title={Wide-bandgap semiconductors: a critical analysis of GaN, SiC, AlGaN, diamond, and Ga2O3 synthesis methods, challenges, and prospective technological innovations},
  author={Yeboah, Luckman Aborah and Abdul Malik, Ayinawu and Oppong, Peter Agyemang and Acheampong, Prince Sarfo and Morgan, Joseph Arko and Addo, Rose Akua Adwubi and Williams Henyo, Boris and Taylor, Stephen Takyi and Zudor, Wolalorm Makafui and Osei-Amponsah, Samuel},
  journal={Intelligent and Sustainable Manufacturing},
  volume={2},
  number={1},
  pages={10011},
  year={2025},
  publisher={SCIEPublish}
}

@article{zhang2022optimal,
  title={Optimal design of GaN HEMT based high efficiency LLC converter},
  author={Zhang, Botao and Zhao, Min and Huang, Pei and Wang, Qi},
  journal={Energy Reports},
  volume={8},
  pages={1181--1190},
  year={2022},
  publisher={Elsevier}
}

@article{han2025thermal,
  title={Thermal management of wide-bandgap power semiconductors: Strategies and challenges in SiC and GaN power devices},
  author={Han, Gyuyeon and Kim, Junseok and Park, Sanghyun and Bae, Wongyu},
  journal={Electronics},
  volume={14},
  number={21},
  pages={4193},
  year={2025},
  publisher={MDPI}
}

@inproceedings{heumesser2023cascode,
  title={Cascode GaN HEMT gate driving analysis},
  author={Heumesser, Vanessa and Lai, Jih-Sheng and Hsieh, Hsin-Che and Hsu, Johnny and Yang, Chih-Yi and Chang, Edward Y and Liu, Ching-Yao and Chieng, Wei-Hua and Hsieh, Yueh-Tsung},
  booktitle={2023 IEEE Workshop on Wide Bandgap Power Devices and Applications in Asia (WiPDA Asia)},
  pages={1--6},
  year={2023},
  organization={IEEE}
}

@article{bau2020cmos,
  title={CMOS active gate driver for closed-loop dv/dt control of GaN transistors},
  author={Bau, Plinio and Cousineau, Marc and Cougo, Bernardo and Richardeau, Fr{\'e}d{\'e}ric and Rouger, Nicolas},
  journal={IEEE Transactions on Power Electronics},
  volume={35},
  number={12},
  pages={13322--13332},
  year={2020},
  publisher={IEEE}
}

@article{bu2020gan,
  title={GaN-based matrix converter design with output filters for motor friendly drive system},
  author={Bu, Hanyoung and Cho, Younghoon},
  journal={Energies},
  volume={13},
  number={4},
  pages={971},
  year={2020},
  publisher={MDPI}
}

@article{lu2025design,
  title={Design and application of high-efficiency gallium nitride (GaN)-based power electronic devices},
  author={Lu, Chen},
  journal={Applied and Computational Engineering},
  volume={153},
  number={1},
  pages={90--95},
  year={2025}
}

@article{hsu2021development,
  title={Development of GaN HEMTs fabricated on silicon, silicon-on-insulator, and engineered substrates and the heterogeneous integration},
  author={Hsu, Lung-Hsing and Lai, Yung-Yu and Tu, Po-Tsung and Langpoklakpam, Catherine and Chang, Ya-Ting and Huang, Yu-Wen and Lee, Wen-Chung and Tzou, An-Jye and Cheng, Yuh-Jen and Lin, Chun-Hsiung and others},
  journal={Micromachines},
  volume={12},
  number={10},
  pages={1159},
  year={2021},
  publisher={MDPI}
}

@article{kaminski2014sic,
  title={SiC and GaN devices--wide bandgap is not all the same},
  author={Kaminski, Nando and Hilt, Oliver},
  journal={IET Circuits, Devices \& Systems},
  volume={8},
  number={3},
  pages={227--236},
  year={2014},
  publisher={Wiley Online Library}
}

@article{dadgar2015sixteen,
  title={Sixteen years GaN on Si},
  author={Dadgar, Armin},
  journal={physica status solidi (b)},
  volume={252},
  number={5},
  pages={1063--1068},
  year={2015},
  publisher={Wiley Online Library}
}

@article{spiridon2021method,
  title={Method for inferring the mechanical strain of GaN-on-Si epitaxial layers using optical profilometry and finite element analysis},
  author={Spiridon, BF and Toon, Miles and Hinz, Alexander and Ghosh, Saptarsi and Fairclough, SM and Guilhabert, BJE and Strain, MJ and Watson, IM and Dawson, Martin D and Wallis, DJ and others},
  journal={Optical Materials Express},
  volume={11},
  number={6},
  pages={1643--1655},
  year={2021},
  publisher={Optical Society of America}
}

@article{jarndal20212,
  title={2-mm-gate-periphery GaN high electron mobility transistor s on SiC and Si substrates: A comparative analysis from a small-signal standpoint},
  author={Jarndal, Anwar and Alim, Mohammad Abdul and Raffo, Antonio and Crupi, Giovanni},
  journal={International Journal of RF and Microwave Computer-Aided Engineering},
  volume={31},
  number={6},
  pages={e22642},
  year={2021},
  publisher={Wiley Online Library}
}

@article{paskova2009gan,
  title={GaN substrates for III-nitride devices},
  author={Paskova, Tanya and Hanser, Drew A and Evans, Keith R},
  journal={Proceedings of the IEEE},
  volume={98},
  number={7},
  pages={1324--1338},
  year={2009},
  publisher={IEEE}
}

@article{huang2024direct,
  title={Direct Growth of Wafer-Scale Self-Separated GaN on Reusable 2D Material Substrates},
  author={Huang, Chang-Hsun and Wu, Chia-Yi and Chou, Yi-Chia},
  journal={Advanced Science},
  volume={11},
  number={41},
  pages={2406126},
  year={2024},
  publisher={Wiley Online Library}
}

@article{liu2021evolution,
  title={The evolution of manufacturing technology for GaN electronic devices},
  author={Liu, An-Chen and Tu, Po-Tsung and Langpoklakpam, Catherine and Huang, Yu-Wen and Chang, Ya-Ting and Tzou, An-Jye and Hsu, Lung-Hsing and Lin, Chun-Hsiung and Kuo, Hao-Chung and Chang, Edward Yi},
  journal={Micromachines},
  volume={12},
  number={7},
  pages={737},
  year={2021},
  publisher={MDPI}
}

@article{zhong2022review,
  title={A review on the GaN-on-Si power electronic devices},
  author={Zhong, Yaozong and Zhang, Jinwei and Wu, Shan and Jia, Lifang and Yang, Xuelin and Liu, Yang and Zhang, Yun and Sun, Qian},
  journal={Fundamental Research},
  volume={2},
  number={3},
  pages={462--475},
  year={2022},
  publisher={Elsevier}
}

@article{lin2020gallium,
  title={Gallium nitride (GaN) high-electron-mobility transistors with thick copper metallization featuring a power density of 8.2 W/mm for Ka-band applications},
  author={Lin, YC and Chen, SH and Lee, PH and Lai, KH and Huang, TJ and Chang, Edward Y and Hsu, Heng-Tung},
  journal={Micromachines},
  volume={11},
  number={2},
  pages={222},
  year={2020},
  publisher={MDPI}
}

@article{jorgensen2018fast,
  title={A fast-switching integrated full-bridge power module based on GaN eHEMT devices},
  author={J{\o}rgensen, Asger Bj{\o}rn and B{\k{e}}czkowski, Szymon and Uhrenfeldt, Christian and Petersen, Niels H{\o}gholt and J{\o}rgensen, S{\o}ren and Munk-Nielsen, Stig},
  journal={IEEE Transactions on Power Electronics},
  volume={34},
  number={3},
  pages={2494--2504},
  year={2018},
  publisher={IEEE}
}

@article{wang2022review,
  title={Review of topside interconnections for wide bandgap power semiconductor packaging},
  author={Wang, Lisheng and Wang, Wenbo and Hueting, Raymond JE and Rietveld, Gert and Ferreira, Jan Abraham},
  journal={IEEE Transactions on Power Electronics},
  volume={38},
  number={1},
  pages={472--490},
  year={2022},
  publisher={IEEE}
}

@article{lumbreras2021effect,
  title={Effect of the heat dissipation system on hard-switching GaN-based power converters for energy conversion},
  author={Lumbreras, David and Vilella, Manel and Zaragoza, Jordi and Berbel, N{\'e}stor and Jord{\`a}, Josep and Collado, Alfonso},
  journal={Energies},
  volume={14},
  number={19},
  pages={6287},
  year={2021},
  publisher={MDPI}
}

@article{calzolaro2022status,
  title={Status of aluminum oxide gate dielectric technology for insulated-gate GaN-based devices},
  author={Calzolaro, Anthony and Mikolajick, Thomas and Wachowiak, Andre},
  journal={Materials},
  volume={15},
  number={3},
  pages={791},
  year={2022},
  publisher={MDPI}
}

@article{manganelli2021strategies,
  title={Strategies for improving the sustainability of data centers via energy mix, energy conservation, and circular energy},
  author={Manganelli, Matteo and Soldati, Alessandro and Martirano, Luigi and Ramakrishna, Seeram},
  journal={Sustainability},
  volume={13},
  number={11},
  pages={6114},
  year={2021},
  publisher={MDPI}
}

@article{yang2022increasing,
  title={Increasing the energy efficiency of a data center based on machine learning},
  author={Yang, Zhen and Du, Jinhong and Lin, Yiting and Du, Zhen and Xia, Li and Zhao, Qianchuan and Guan, Xiaohong},
  journal={Journal of Industrial Ecology},
  volume={26},
  number={1},
  pages={323--335},
  year={2022},
  publisher={Wiley Online Library}
}

@article{zhang2022prediction,
  title={Prediction of overall energy consumption of data centers in different locations},
  author={Zhang, Yiliu and Liu, Jie},
  journal={Sensors},
  volume={22},
  number={10},
  pages={3704},
  year={2022},
  publisher={MDPI}
}

@article{mondal2023geeco,
  title={GEECO: Green data centers for energy optimization and carbon footprint reduction},
  author={Mondal, Sudipto and Faruk, Fashat Bin and Rajbongshi, Dibosh and Efaz, Mohammad Masum Khondhoker and Islam, Md Motaharul},
  journal={Sustainability},
  volume={15},
  number={21},
  pages={15249},
  year={2023},
  publisher={MDPI}
}

@article{fieni2025x,
  title={$ x $ PUE: Extending Power Usage Effectiveness Metrics for Cloud Infrastructures},
  author={Fieni, Guillaume and Rouvoy, Romain and Seinturier, Lionel},
  journal={IEEE Transactions on Sustainable Computing},
  year={2025},
  publisher={IEEE}
}

@article{liu2025carbon,
  title={Carbon emission modeling for high-performance computing-based AI in new power systems with large-scale renewable energy integration},
  author={Liu, Haoyang and Zhai, Jiangtao},
  journal={Processes},
  volume={13},
  number={2},
  pages={595},
  year={2025},
  publisher={MDPI}
}

@inproceedings{gupta2021chasing,
  title={Chasing carbon: The elusive environmental footprint of computing},
  author={Gupta, Udit and Kim, Young Geun and Lee, Sylvia and Tse, Jordan and Lee, Hsien-Hsin S and Wei, Gu-Yeon and Brooks, David and Wu, Carole-Jean},
  booktitle={2021 IEEE International Symposium on High-Performance Computer Architecture (HPCA)},
  pages={854--867},
  year={2021},
  organization={IEEE}
}

@inproceedings{li2023toward,
  title={Toward sustainable hpc: Carbon footprint estimation and environmental implications of hpc systems},
  author={Li, Baolin and Basu Roy, Rohan and Wang, Daniel and Samsi, Siddharth and Gadepally, Vijay and Tiwari, Devesh},
  booktitle={Proceedings of the international conference for high performance computing, networking, storage and analysis},
  pages={1--15},
  year={2023}
}

@article{monch2023highly,
  title={How highly efficient power electronics transfers high electrocaloric material performance to heat pump systems: S. M{\"o}nch et al.},
  author={M{\"o}nch, Stefan and Reiner, Richard and Waltereit, Patrick and Basler, Michael and Quay, R{\"u}diger and Gebhardt, Sylvia and Molin, Christian and Bach, David and Binninger, Roland and Bartholom{\'e}, Kilian},
  journal={MRS advances},
  volume={8},
  number={15},
  pages={787--796},
  year={2023},
  publisher={Springer}
}

@article{sesotyo2025evaluating,
  title={Evaluating of DC-DC Buck-Boost Converter implementation for Integrated Solar Photovoltaic and Thermoelectric Cooler System},
  author={Sesotyo, Priyo Adi and Cahyono, Taufik Dwi and Sadewa, Ery and others},
  journal={International Journal of Engineering Continuity},
  volume={4},
  number={1},
  pages={140--172},
  year={2025}
}

@article{vikhor2024modeling,
  title={Modeling of thermoelectric converter characteristics: Lecture at the Summer Thermoelectric School, June 30, 2024, Krakow, Poland},
  author={Vikhor, LM},
  journal={Journal of Thermoelectricity},
  number={3},
  pages={5--22},
  year={2024}
}

@article{le2019techno,
  title={Techno-economic assessment of cascade air-to-water heat pump retrofitted into residential buildings using experimentally validated simulations},
  author={Le, Khoa Xuan and Huang, Ming Jun and Shah, Nikhilkumar N and Wilson, Christopher and Mac Artain, Paul and Byrne, Raymond and Hewitt, Neil J},
  journal={Applied Energy},
  volume={250},
  pages={633--652},
  year={2019},
  publisher={Elsevier}
}

@techreport{pilawa2024extreme,
  title={Extreme Efficiency 240 Vac to Load Data Center Power Delivery Topologies and Control},
  author={Pilawa-Podgurski, Robert},
  year={2024},
  institution={University of California, Berkeley, CA (United States)}
}

@article{lee2022design,
  title={Design of a GaN totem-pole PFC converter using DC-link voltage control strategy for data center applications},
  author={Lee, Jia-You and Chen, Jheng-Hung and Lo, Kuo-Yuan},
  journal={IEEE Access},
  volume={10},
  pages={50278--50287},
  year={2022},
  publisher={IEEE}
}

@article{sun2021mitigation,
  title={Mitigation of current distortion for GaN-based CRM totem-pole PFC rectifier with ZVS control},
  author={Sun, Jingjing and Gui, Handong and Li, Jie and Huang, Xingxuan and Strain, Nathan and Costinett, Daniel J and Tolbert, Leon M},
  journal={IEEE Open Journal of Power Electronics},
  volume={2},
  pages={290--303},
  year={2021},
  publisher={IEEE}
}

@article{huang2023cooling,
  title={Cooling technologies for internet data center in China: Principle, energy efficiency, and applications},
  author={Huang, Xiaofei and Yan, Junwei and Zhou, Xuan and Wu, Yixin and Hu, Shichen},
  journal={Energies},
  volume={16},
  number={20},
  pages={7158},
  year={2023},
  publisher={MDPI}
}

@article{chang2024optimization,
  title={Optimization control strategies and evaluation metrics of cooling systems in data centers: a review},
  author={Chang, Qiankun and Huang, Yuanfeng and Liu, Kaiyan and Xu, Xin and Zhao, Yaohua and Pan, Song},
  journal={Sustainability},
  volume={16},
  number={16},
  pages={7222},
  year={2024},
  publisher={MDPI}
}

@article{ishraq2024design,
  title={Design of a 2.5 kw four-level interleaved flying capacitor multilevel totem-pole pfc converter with ac-side passive volume optimization},
  author={Ishraq, Naveed and Mallik, Ayan},
  journal={IEEE Open Journal of Power Electronics},
  volume={5},
  pages={214--231},
  year={2024},
  publisher={IEEE}
}

@article{okilly2022design,
  title={Design and Fabrication of an Isolated Two-Stage AC--DC Power Supply with a 99.50\% PF and ZVS for High-Power Density Industrial Applications},
  author={Okilly, Ahmed H and Baek, Jeihoon},
  journal={Electronics},
  volume={11},
  number={12},
  pages={1898},
  year={2022},
  publisher={MDPI}
}

@article{chrysostomou2021multicell,
  title={Multicell power supplies for improved energy efficiency in the information and communications technology infrastructures},
  author={Chrysostomou, Michael and Christofides, Nicholas and Ioannou, Stelios and Polycarpou, Alexis},
  journal={Energies},
  volume={14},
  number={21},
  pages={7038},
  year={2021},
  publisher={MDPI}
}

@article{reusch2018system,
  title={System optimization of a high power density non-isolated intermediate bus converter for 48 V server applications},
  author={Reusch, David and Biswas, Suvankar and Zhang, Yuanzhe},
  journal={IEEE Transactions on Industry Applications},
  volume={55},
  number={2},
  pages={1619--1627},
  year={2018},
  publisher={IEEE}
}

@article{ahmed2020two,
  title={Two-stage 48-V VRM with intermediate bus voltage optimization for data centers},
  author={Ahmed, Mohamed H and Lee, Fred C and Li, Qiang},
  journal={IEEE Journal of Emerging and Selected Topics in Power Electronics},
  volume={9},
  number={1},
  pages={702--715},
  year={2020},
  publisher={IEEE}
}

@article{wang20243,
  title={A 3 kw gan hemt based three-phase converter achieving a switching frequency of 300 khz and an efficiency of 97.06\%},
  author={Wang, Zihao and Ye, Fei and Duan, Shunshuai and Yuan, Xibo and Zuo, Dongsheng and Zhang, Yonglei and Wang, Kai and Li, Yan},
  journal={IEEE Access},
  volume={12},
  pages={116442--116456},
  year={2024},
  publisher={IEEE}
}

@article{li2023survey,
  title={A survey of conductive and radiated EMI reduction techniques in power electronics converters across wide-bandgap devices},
  author={Li, Chentao and Ma, Qishuang and Tong, Yajing and Wang, Jinsong and Xu, Ping},
  journal={IET Power Electronics},
  volume={16},
  number={13},
  pages={2121--2137},
  year={2023},
  publisher={Wiley Online Library}
}

@article{lee2016application,
  title={Application of GaN devices for 1 kW server power supply with integrated magnetics},
  author={Lee, Fred C and Li, Qiang and Liu, Zhengyang and Yang, Yuchen and Fei, Chao and Mu, Mingkai},
  journal={CPSS Transactions on power electronics and applications},
  volume={1},
  number={1},
  pages={3--12},
  year={2016},
  publisher={CPSS}
}

@article{jia2024mitigating,
  title={Mitigating EMI noise in propagation paths: Review of parasitic and coupling effects in power electronic packages, filters, and systems},
  author={Jia, Niu and Xue, Lingxiao and Cui, Han},
  journal={IEEE Open Journal of Power Electronics},
  volume={5},
  pages={352--368},
  year={2024},
  publisher={IEEE}
}

@article{rothmund201899,
  title={99\% efficient 10 kV SiC-based 7 kV/400 V DC transformer for future data centers},
  author={Rothmund, Daniel and Guillod, Thomas and Bortis, Dominik and Kolar, Johann W},
  journal={IEEE Journal of Emerging and Selected Topics in Power Electronics},
  volume={7},
  number={2},
  pages={753--767},
  year={2018},
  publisher={IEEE}
}

@article{prajapati2023leveraging,
  title={Leveraging GaN for DC-DC power modules for efficient EVs: A review},
  author={Prajapati, Paramanand and Balamurugan, S},
  journal={IEEE Access},
  volume={11},
  pages={95874--95888},
  year={2023},
  publisher={IEEE}
}

@inproceedings{sun2019research,
  title={Research of PCB parasitic inductance in the GaN transistor power loop},
  author={Sun, Bainan and Zhang, Zhe and Andersen, Michael AE},
  booktitle={2019 IEEE Workshop on Wide Bandgap Power Devices and Applications in Asia (WiPDA Asia)},
  pages={1--5},
  year={2019},
  organization={IEEE}
}

@article{derkacz2022emi,
  title={EMI mitigation of GaN power inverter leg by local shielding techniques},
  author={Derkacz, Pawel B and Schanen, Jean-Luc and Jeannin, Pierre-Olivier and Chrzan, Piotr J and Musznicki, Piotr and Petit, Mickael},
  journal={IEEE Transactions on Power Electronics},
  volume={37},
  number={10},
  pages={11996--12004},
  year={2022},
  publisher={IEEE}
}

@article{lavrivc2025challenges,
  title={Challenges for large-scale deployment of WBG in power electronics},
  author={Lavri{\v{c}}, Henrik and Zajec, Peter and Drobni{\v{c}}, Klemen and Rihar, Andra{\v{z}} and Ambro{\v{z}}i{\v{c}}, Vanja and Von{\v{c}}ina, Danjel and Nemec, Mitja},
  journal={Informacije MIDEM},
  volume={55},
  number={1},
  pages={3--24},
  year={2025}
}

@article{qin2023thermal,
  title={Thermal management and packaging of wide and ultra-wide bandgap power devices: a review and perspective},
  author={Qin, Yuan and Albano, Benjamin and Spencer, Joseph and Lundh, James Spencer and Wang, Boyan and Buttay, Cyril and Tadjer, Marko and DiMarino, Christina and Zhang, Yuhao},
  journal={Journal of physics D: applied physics},
  volume={56},
  number={9},
  pages={093001},
  year={2023},
  publisher={IOP Publishing}
}

@article{mukherjee2021challenges,
  title={Challenges and perspectives for vertical GaN-on-Si trench MOS reliability: From leakage current analysis to gate stack optimization},
  author={Mukherjee, Kalparupa and De Santi, Carlo and Borga, Matteo and Geens, Karen and You, Shuzhen and Bakeroot, Benoit and Decoutere, Stefaan and Diehle, Patrick and H{\"u}bner, Susanne and Altmann, Frank and others},
  journal={Materials},
  volume={14},
  number={9},
  pages={2316},
  year={2021},
  publisher={MDPI}
}

@article{wang2024advanced,
  title={Advanced packaging technology of GaN HEMT module for high-power and high-frequency applications: a review},
  author={Wang, Meiyu and Gao, Peng and Shi, Fangmin and Hu, Weibo and Wang, Xudong and Yan, Haidong and Mei, Yunhui},
  journal={IEEE Transactions on Components, Packaging and Manufacturing Technology},
  volume={14},
  number={9},
  pages={1537--1550},
  year={2024},
  publisher={IEEE}
}

@article{wohrle2023power,
  title={Power Module Design for GaN Transistors Enabling High Switching Speed in Multi-Kilowatt Applications},
  author={W{\"o}hrle, Dennis and Burger, Bruno and Ambacher, Oliver},
  journal={Energy Technology},
  volume={11},
  number={12},
  pages={2300460},
  year={2023},
  publisher={Wiley Online Library}
}

@article{sun2024thermal,
  title={Thermal cycling characterization of an integrated low-inductance GaN eHEMT power module},
  author={Sun, Zhongchao and Takahashi, Masaki and Guo, Wendi and Munk-Nielsen, Stig and J{\o}rgensen, Asger Bj{\o}rn},
  journal={Microelectronics Reliability},
  volume={161},
  pages={115482},
  year={2024},
  publisher={Elsevier}
}

@article{wesselkaemper2025enhancing,
  title={Enhancing supply resilience for critical materials: Case study of gallium supply in the United States},
  author={Wesselkaemper, Jannis and Newkirk, Alex C and Hendrickson, Thomas P and Helal, Nadiyah and Rao, Prakash and Smith, Sarah J and Haddad, Andrew Z},
  journal={Resources, Conservation and Recycling},
  volume={222},
  pages={108436},
  year={2025},
  publisher={Elsevier}
}

@article{Kozak2023GaNReview,
  author  = {Kozak, Joseph Peter and Zhang, Ruizhe and Porter, Matthew and
             Song, Qihao and Liu, Jingcun and Wang, Bixuan and Wang, Rudy and
             Saito, Wataru and Zhang, Yuhao},
  title   = {Stability, Reliability, and Robustness of {GaN} Power Devices:
             A Review},
  journal = {IEEE Transactions on Power Electronics},
  year    = {2023},
  volume  = {38},
  number  = {7},
  pages   = {8442--8471},
  month   = jul,
  doi     = {10.1109/TPEL.2023.3266365}
}

@article{Rauf2025Investigation,
  author  = {Rauf, Fawad and Tayyab, Muhammad Farhan and Mouhoubi, Samir and
             Heldwein, Marcelo Lobo and Curatola, Gilberto},
  title   = {Investigation of the Long-Term Dynamic {$R_{\mathrm{DS(on)}}$}
             Variation and Dynamic High Temperature Operating Life Test
             Robustness of {Schottky}-Gate and Ohmic-Gate {GaN HEMT} with
             Comparable Stress Conditions},
  journal = {Microelectronics Reliability},
  year    = {2025},
  volume  = {168},
  pages   = {115708},
  month   = may,
  doi     = {10.1016/j.microrel.2025.115708}
}

@article{Tayyab2022Dynamic,
  author  = {Tayyab, Muhammad Farhan and Basler, Thomas},
  title   = {Dynamic High Temperature Operating Life Test Methodology for
             Long-Term Switching Reliability of {GaN} Power Devices},
  journal = {Microelectronics Reliability},
  year    = {2022},
  volume  = {138},
  pages   = {114613},
  month   = nov,
  doi     = {10.1016/j.microrel.2022.114613}
}

@article{Fan2024Dynamic,
  author  = {Fan, Chen and Zhang, Haitao and Liu, Huipeng and Pan, Xiaofei and
             Yan, Su and Chen, Hongliang and Guo, Wei and Cai, Lin and Wei, Shuhua},
  title   = {A Study on the Dynamic Switching Characteristics of {p-GaN HEMT}
             Power Devices},
  journal = {Micromachines},
  year    = {2024},
  volume  = {15},
  number  = {8},
  pages   = {993},
  month   = jul,
  doi     = {10.3390/mi15080993}
}

@techreport{JEDEC2021JEP180,
  author      = {{JEDEC Solid State Technology Association}},
  title       = {Guideline for Switching Reliability Evaluation Procedures for
                 Gallium Nitride Power Conversion Devices},
  institution = {JEDEC Solid State Technology Association},
  type        = {JEDEC Publication},
  number      = {JEP180.01},
  year        = {2021},
  month       = jan
}

\end{document}